\documentclass[format=acmsmall, review=false]{acmart}
\usepackage{acm-ec-26}
\usepackage{booktabs} 
\usepackage[ruled]{algorithm2e} 

\SetAlFnt{\small}
\SetAlCapFnt{\small}
\SetAlCapNameFnt{\small}
\SetAlCapHSkip{0pt}
\IncMargin{-\parindent}

\setcitestyle{authoryear}

\usepackage{tikz}
\usetikzlibrary{matrix}
\usetikzlibrary{positioning}

\usepackage{subcaption}
\usepackage{graphicx}
\usepackage{multirow}
\newtheoremstyle{theorem}  
 {\topsep}                
 {\topsep}               
 {\itshape}              
 {}                      
 {\bfseries}             
 {.}                     
 { }                    
 {}                     

\theoremstyle{theorem}
\newtheorem{theorem}{Theorem}
\newtheorem{proposition}{Proposition}
\newtheorem{assumption}{Assumption}
\newtheorem{lemma}{Lemma}

\newtheorem{remark}{Remark}

\title[Experimental Designs for Multi-Item Multi-Period Inventory Control]{Experimental Designs for Multi-Item Multi-Period Inventory Control}

\author{Xinqi Chen}
\affiliation{%
  \institution{University of California, Berkeley}
  \city{Berkeley}
  \state{CA}
  \country{USA}}
\email{xinqi\_chen@berkeley.edu}

\author{Xingyu Bai}
\affiliation{%
  \institution{The Hong Kong Polytechnic University}
  \city{Hong Kong}
  \country{Hong Kong SAR, China}}
\email{xingyu.bai@polyu.edu.hk}

\author{Zeyu Zheng}
\affiliation{%
  \institution{University of California, Berkeley}
  \city{Berkeley}
  \state{CA}
  \country{USA}}
\email{zyzheng@berkeley.edu}

\author{Nian Si}
\affiliation{%
  \institution{The Hong Kong University of Science and Technology}
  \city{Hong Kong}
  \country{Hong Kong SAR, China}}
\email{niansi@ust.hk}



\begin{abstract}
Randomized experiments, or A/B testing, are the gold standard for evaluating interventions, yet they remain underutilized in inventory management. This study addresses this gap by analyzing A/B testing strategies in multi-item, multi-period inventory systems with lost sales and capacity constraints. We examine two canonical experimental designs—switchback experiments and item-level randomization—and show that both suffer from systematic bias due to interference: temporal carryover in switchbacks and cannibalization across items under capacity constraints. Under mild conditions, we characterize the direction of this bias in different scenarios. Motivated by two-sided randomization, we propose a pairwise design over items and time and analyze its bias properties. Controlled stochastic simulations verify the theoretical predictions, and trace-driven experiments on real-world fresh-retail data show that the same mechanisms persist in realistic environments with stockout substitution.
\end{abstract}

\begin{document}

\def\EE{\mathbb{E}}
\def\PP{\mathbb{P}}
\def\RR{\mathbb{R}}
\def\Rcal{\mathcal{R}}
\def\Ycal{\mathcal{Y}}

\def\A{\mathcal{A}}
\def\W{\mathbf{W}}
\def\X{\mathbf{X}}
\def\I{\mathbf{I}}
\def\O{\mathbf{O}}
\def\S{\mathbf{S}}

\def\PP{\mathbb{P}}
\def\RR{\mathbb{R}}
\def\YY{\mathbb{Y}}

\def\Wcal{\mathcal{W}}
\def\Ocal{\mathcal{O}}

\def\bw{\boldsymbol{w}}
\def\bk{\boldsymbol{k}}
\def\bv{\boldsymbol{v}}
\def\bu{\boldsymbol{u}}
\def\bs{\boldsymbol{s}}
\def\bw{\boldsymbol{w}}
\def\bd{\boldsymbol{d}}

\def\S{\mathbf{S}}
\def\Y{\mathbf{Y}}
\def\s{\mathbf{s}}
\def\D{\mathbf{D}}

 \newcommand{\1}{\mathbf 1} 
\newcommand{\Var}{\operatorname{Var}}
\newcommand{\Normal}{\mathcal N}

\newcommand{\clip}[1]{\left[#1\right]_0^1}

\newif\ifshowreviewcomments
\showreviewcommentstrue

\long\def\DraftReviewComment#1{%
  \ifshowreviewcomments
    \par\noindent
    \begingroup
    \color{blue}
    \fbox{\begin{minipage}{0.95\linewidth}
    \textbf{Review comment.} #1
    \end{minipage}}
    \endgroup
    \par
  \fi
}

\begin{titlepage}

\maketitle

\vspace{1cm}
\setcounter{tocdepth}{2} 

\end{titlepage}

\section{Introduction}

Across modern digital platforms, pharmaceutical innovation, and empirical economics, researchers rely on controlled experiments, such as A/B testing, clinical trials, and randomized controlled trials (RCTs), to rigorously evaluate the effectiveness of new strategies or interventions.
The fundamental principle of traditional randomized experiments involves randomly assigning experimental units to treatment and control groups, then inferring the effect of intervention by measuring the average performance difference between groups on metrics of interest. In practical applications, decision makers employ A/B testing to evaluate whether implementing a new intervention (\textit{treatment}) across the entire platform would yield better performance compared to the existing approach (\textit{control}), referred to as Global Treatment Effect (GTE). 

On large online platforms, thousands of experiments are conducted daily to assess the impact of changes, ranging from minor adjustments to major overhauls. Extensive literature exists on bias analysis across different experimental designs and platforms, including examples from room-sharing platforms \citep{johari2022experimental}, advertising platforms \citep{basse2016randomization,liao2024interference}, recommendation systems  \citep{si2023tackling, farias2023correcting, zhan2024estimating}, ride-sharing platforms \citep{bright2024reducing}, and gaming platforms \citep{weng2024experimental}.

While the technology industry has widely embraced A/B testing, such data-driven methods remain underexplored and underutilized in other industries, such as inventory management. However, as Prof. Thomke stated in his book 
\textit{Experimentation Works: The Surprising Power of Business Experiments}  \citep{thomke2020experimentation}:
\begin{quote}
``Large-scale, controlled experimentation would revolutionize the way all companies operate their businesses and how managers make decisions."
\end{quote}
We also anticipate that A/B testing will increasingly be adopted in those industries as well. In particular, industries such as inventory management have witnessed a prominent trend to become more and more data-oriented, providing an advantageous position to effectively implement A/B tests. 

This paper studies A/B testing for inventory management. In recent years, a growing body of work has proposed automatic, data-driven inventory control methods \citep{shi2016nonparametric,lyu2024minibatch,xie2024vc}. While these methods exhibit strong theoretical guarantees under stylized assumptions, how to reliably evaluate their practical performance in real-world environments—where such assumptions are frequently violated—remains largely underexplored. The dominant evaluation paradigm relies on simulation: practitioners estimate demand distributions from historical data and construct behavioral models to capture customer responses after stockouts. However, this approach has fundamental limitations: inventory data are censored, so post-stockout demand is unobservable yet often decisive for assessing policy performance; post-stockout customer behavior (substitution, abandonment, delayed purchase) is complex and hard to model robustly; and high-fidelity, large-scale simulation is computationally and operationally costly in multi-item, multi-period settings. These challenges cast doubt on the reliability and scalability of simulation-based evaluation.

Motivated by these limitations, this study investigates A/B testing strategies for evaluating inventory control policies and provides guidance for experimental design across different operational regimes.
Specifically, we focus on A/B testing different inventory policies in  multi-item, multi-period systems  with lost sales and a warehouse capacity constraint \citep{snyder2019fundamentals}. We analyze two canonical experimental designs: switchback experiments \citep{cochran1941double,bojinov2023design,xiong2023data} and item-level randomization. 
In switchback experiments, the treatment assignment for all experimental items randomly alternates between the control group and the treatment group over time. In contrast, item-level randomization involves randomly assigning items to treatment and control groups, with their treatment assignments remaining fixed throughout the experiment.

However, both the aforementioned experimental designs have limitations due to bias caused by interference. In switchback experiments, temporal interference or carryover effects are present, meaning that a treatment not only impacts the outcome of the current period but also influences the system’s state and subsequent outcomes in future periods. For item-level randomization, cannibalization effects may occur due to capacity constraints, where the treatment assigned to one item affects the outcomes of other items. 
These scenarios violate the Stable Unit Treatment Value Assumption (SUTVA) in causal inference \citep{imbens2015causal}.  Previous studies showed that the bias caused by interference can be as large as the GTE itself \citep{blake2014marketplace,fradkin2019simulation,holtz2020reducing}.

In this work, we examine the bias of the naive inverse probability weighting (IPW) estimator under different experimental designs, including switchback experiments and item-level randomized experiments. Our contributions are summarized as follows:

\begin{enumerate}
\item We investigate A/B testing strategies in multi-item, multi-period inventory systems with lost sales and capacity constraints. The treatment and control differ in their demand forecasters. When the primary difference between the two forecasters lies in their means, we show that switchback designs tend to systematically underestimate the overall treatment effect, whereas item-level randomization tends to systematically overestimate it. In contrast, when the forecasters mainly differ in their forecast-error dispersion, switchback designs typically overestimate the treatment effect, while item-level randomization is asymptotically unbiased.

\item Motivated by \citet{johari2022experimental} and \citet{bajari2023experimental}, we propose a pairwise randomized experiment that treats products and time periods as two-sided factors and analyze its bias properties. We find that when the treatment and control primarily differ in forecasting means, pairwise randomization yields lower bias than item-level randomization. Conversely, when the primary difference lies in forecast-error dispersion, pairwise randomization exhibits higher asymptotic bias than item-level randomization.

\item We conduct two complementary numerical studies. Controlled stochastic simulations match the uniform-demand assumptions in our theory and verify the predicted signs. Trace-driven experiments using FreshRetailNet-50K \citep{wang2025freshretailnet} then evaluate the same mechanisms with recovered real demand traces, learned point forecasts, and stockout substitution. The results provide practical recommendations for selecting experimental designs under different operational settings.
\end{enumerate}

The rest of the paper is organized as follows. In Section \ref{section1.1} we review related work on A/B testing and multi-item inventory control. Section \ref{section:model}  specifies the model of multi-item multi-period inventory control, and the inventory policy we use. In Section \ref{section:design}, we introduce the experimental designs and the treatment effect estimator.  In Section~\ref{sec:analysis}, we study A/B testing of inventory policies induced by different demand estimates and characterize the sign of the bias in the estimated global treatment effect under different experimental designs. 
In Section~\ref{sec:numerical}, we first conduct controlled stochastic simulations that directly match the theory and then report trace-driven experiments using real-world fresh-retail data.
Finally, we conclude with several practical implications and future work in Section \ref{section:conclusion}.

\subsection{Related Works}\label{section1.1}
\textbf{Experimental designs and bias analysis.} There is extensive research on experimental designs in two-sided marketplaces, such as \cite{pouget2019variance, holtz2020reducing, johari2022experimental, li2022interference, bajari2023experimental, dhaouadi2023price, harshaw2023design, bright2024reducing, masoero2024multiple, zhan2024estimating}. While our setting is also two-sided (product and temporal sides), it is ``asymmetric,'' unlike the ``symmetric'' demand and supply sides of traditional marketplaces. \citet{masoero2023efficient} adopts a similar view; we differ by systematically analyzing designs within the context of inventory control.

Switchback experiments and temporal interference have been also widely studied in the literature; see, for example, \citet{glynn2020adaptive,hu2022switchback,bojinov2023design,xiong2023data,ni2023design,li2023experimenting,han2024population}. And some researchers are using reinforcement learning techniques to tackle this type of interference \citep{farias2022markovian,shi2023dynamic,chen2024experimenting}. Our work specifically focuses on inventory management, enabling us to characterize carryover effects and perform detailed bias analysis. Additionally, some studies address scenarios with irreversible treatment conditions, where items assigned to the treatment group cannot revert to the control group. In such cases, staggered rollout designs are used, gradually increasing the proportion of individuals in the treatment group. \citet{han2023detecting} and \citet{boyarsky2023modeling} developed statistical tests for interference in A/B testing with increasing allocation, while \citet{xiong2024optimal} analyzed this design and established optimality conditions for both non-adaptive and adaptive settings.

Other types of experimental designs and interference structures have also been  studied in the literature. These include clustering randomization and network interference \citep{ugander2013graph,aronow2017estimating,pouget2019variance,jagadeesan2020designs,doudchenko2020causal,yu2022estimating,leung2022rate,harshaw2023design,jia2023clustered,candogan2024correlated,shirani2024causal}, experimental designs in auctions \citep{basse2016randomization,chawla2016b,liao2023statistical}, experiments under non-stationarities \citep{wu2022non, simchi2023non,wu2024nonstationary} and adaptive experimental designs \citep{kasy2021adaptive,qin2022adaptivity}. For further details, we refer readers to the excellent tutorial by \citet{zhao2024experimental}.

For A/B testing in inventory systems, recently \citep{chen2025b} shows that product stockouts create interference across treatment and control groups, causing standard RCT estimators to overestimate treatment effects. Their analysis focuses on bias induced by stockout-driven substitution, whereas our setting highlights how improvements in demand forecasting propagate through inventory decisions and interact with experimental design to generate bias.

\textbf{Multi-item inventory models.} The research of inventory
models with multiple items and capacity constraints dates back to $1960$s.
\cite{veinott1965optimal} and \cite{ignall1969optimality} considered the multi-period model
with stationary demand distributions, and focused on conditions under which a myopic multi-item base-stock policy is optimal. 
\cite{nahmias1984efficient} and \cite{erlebacher2000optimal}
investigated the multi-item newsvendor problem with capacity constraints, introducing both optimal and heuristic methods for solving it.
\cite{beyer2001stochastic,beyer2002average}
considered linear ordering costs and convex backlogging costs, and demonstrated that the base-stock policy is optimal in terms of infinite-horizon discounted costs and long-term average costs, respectively. In addition to a single capacity constraint, \cite{ben1993constrained,lau1995multi,decroix1998optimal,downs2001managing,niederhoff2007using,abdel2007quadratic,zhang2010multi} considered multiple constraints. Readers are referred to \cite{turken2012multi} for a detailed review. More recently, \cite{shi2016nonparametric} and \cite{guo2024online} studied the multi-item lost-sales inventory system with a warehouse
capacity constraint under the lens of demand learning from censored data, and
\cite{federgruen2023multi,federgruen2023scalable}
extended this line of
research by incorporating additional chance constraints and proposing asymptotically optimal policies with a large number of items.

\textbf{Learning algorithms for inventory control.} Recent research has focused on developing data-driven learning algorithms to estimate (possibly censored) demand functions. A recent paper by \citet{xie2024vc} provides a theoretical foundation for this learning problem. 
Learning algorithms in inventory control typically involve Bayesian learning approach \citep{lariviere1999stalking,chen2008dynamic,chu2008solving, araman2009dynamic, wang2017bayesian} where beliefs about the unknown parameters are updated sequentially and optimized with respect to the posterior distribution. More recent work integrates prediction and optimization more tightly, learning decision rules directly from data rather than relying on a separate parametric estimation step \citep{bertsimas2020predictive,elmachtoub2022smart}.

For nonparametric demand learning,  one substantial line of work develops stochastic gradient descent (SGD)–based online algorithms with regret guarantees for inventory systems, including settings with lead times and other model variations \citep{burnetas2000adaptive,huh2009nonparametric,agrawal2019learning,zhang2020closing,chen2020optimal,yuan2021marrying,ding2024feature,yang2024nonparametric,lyu2024minibatch,shi2016nonparametric}. Another stream adopts sample average approximation (SAA) methods for data-driven decision making \citep{kleywegt2002sample,besbes2013implications,levi2015data,lin2022data,fan2022sample,qin2023sailing}. More recent advances explore UCB-type and mirror-descent frameworks for high-dimensional settings \citep{lyu2024ucb,guo2024online}. While these methods offer strong theoretical guarantees, their practical performance has been less systematically evaluated. Our work addresses this gap by developing testing procedures that directly assess real-world performance.

\section{Model Formulation}\label{section:model}

{We consider a periodic-review inventory system of $N$ different items with a shared warehouse capacity $B$ over a planning horizon of $T$ periods.
We consider a lost-sales inventory model with zero replenishment lead time; that is, unmet demand is lost. 
The zero lead time assumption is made for tractability, as the lost-sales problem with positive lead times is known to be notoriously complex, even for a single product (see \cite{zipkin2008structure}).
The true demand distributions are unknown, but the decision maker can have estimates of those distributions (see Section \ref{sec:estimation}), and make inventory decisions accordingly.
}

\subsection{System Dynamics}\label{sec:dynamics}

We denote the random demand vector in period $t\in[T]$ by $\D_t=(D_{1,t},...,D_{N,t})$, where $D_{n,t}$ represents the demand of item $n\in[N]$ in period $t$.
The cumulative distribution function (CDF) of $D_{n,t}$ is denoted by $F_{n,t}$.
In each period $t\in[T]$, the event sequence is given as follows:

    \begin{enumerate}
        \item The decision maker observes the initial inventory level $\I_t = (I_{1,t}, \dots, I_{N,t})$. 
        {\item The decision maker places an order $\O_t=(O_{1,t}, \dots, O_{N,t})$,
        which is  delivered immediately, resulting in the updated inventory level $\S_t = \I_t + \O_t$, where $\S_t=(S_{1,t}, \dots, S_{N,t})$. 
        The total on-hand inventory level must satisfy the warehouse capacity constraint 
        $
            \sum^N_{n=1}S_{n,t}\le B.
        $ 
        \item The random demand $\D_t$ is realized and satisfied by the on-hand inventory to the maximum extent (we use the same notation $\D_t$ for both the random vector and its realization). Any unsatisfied demand is lost.
        }
        \item 
        The profit in this period is  $R_{t} =\sum_{n=1}^N R_{n,t}$, where 
$$R_{n,t} = b_n \min(I_{n,t} +O_{n,t} ,D_{n,t}) - c_n O_{n,t} - h_n(I_{n,t} +O_{n,t} -D_{n,t})^+  ,$$
and $b_n$, $c_n$ and $h_n$ denote the unit selling price, ordering cost, and holding cost for item $n\in[N]$, respectively. Throughout, we assume $b_n>c_n\ge 0$ and $h_n\ge 0$ for all $n$. We write
\[
m_n:=b_n-c_n,\qquad M_n:=b_n-c_n+h_n,\qquad u_n:=\frac{m_n}{M_n},\qquad v_n:=\frac{1}{M_n}.
\]
We further define $R_{n,t}^+(s,d) := m_n s - M_n(s-d)^+$, then $R_{n,t}$ can also be expressed as
 \begin{align*}
     R_{n,t} = R_{n,t}^+(I_{n,t} + O_{n,t},D_{n,t}) - c_n(I_{n,t} + O_{n,t} - D_{n,t})^+ + c_nI_{n,t},\ \forall n \in [N],\ \forall t \in [T-1].
 \end{align*}
 \item The initial inventory for item $n\in[N]$ in the next period is \(     I_{n,t+1}  = (I_{n,t} + O_{n,t} - D_{n,t})^+.
\)
    \end{enumerate}

At the end of period $T$, a salvage value of $c_nI_{n,T+1} = c_n(I_{n,T} + O_{n,T} - D_{n,T})^+$ is generated  for each item $n\in[N]$. For convenience, this term is included in $R_{n,T}$, i.e., 
 \begin{align*}
     R_{n,T} = R_{n,T}^+(I_{n,T} + O_{n,T},D_{n,T}) + c_nI_{n,T},\ \forall n \in [N].
 \end{align*}
The initial inventory levels are zero, i.e., $I_{n,1}=0$ for any $n\in[N]$.

\subsection{Clairvoyant Optimal Inventory Policy}\label{sec:myopic}
According to our definitions of the reward and salvage value, the total expected reward is given by
\begin{align}
\EE \left[ \sum_{n=1}^N \sum_{t=1}^T R_{n,t}  \right]&=\sum_{n=1}^N\left\{  \EE\left[ \sum_{t=1}^T \left(R^+_{n,t}(S_{n,t},D_{n,t}) - c_n I_{n,{t+1}} + c_n I_{n,t} \right) + c_n I_{n,T+1}\right]\right\} \notag \\
&=\sum_{n=1}^N c_n I_{n,1} +\sum_{t=1}^T\EE\left[ \sum_{n=1}^N   R^+_{n,t}(S_{n,t},D_{n,t})  \right].
\label{eq:inentory_equi}
\end{align}

From Equation \eqref{eq:inentory_equi}, a myopic policy that optimizes $\EE\left[ \sum_{n=1}^N   R^+_{n,t}(S_{n,t},D_{n,t})  \right]$ in each period is globally optimal under stationary demand (see, e.g., \cite{ignall1969optimality}; \cite{shi2016nonparametric}). In non-stationary environments, a closed-form global optimum is generally intractable, and the curse of dimensionality worsens this as $N$ and $T$ grow. Following standard approximation approaches (e.g., \cite{chand2002forecast}; \cite{levi2007approximation}), we therefore adopt a rolling-horizon policy with a single-period lookahead, the Myopic Clairvoyant Policy. 
Specifically, given the known demand distribution $\{F_{n,t}\}_{n \in [N]}$ and an initial inventory vector $\I_t=(I_{1,t},\dots,I_{N,t})$ in period $t\in[T]$, the myopic optimization problem for that period is formulated as
\begin{subequations}
\begin{align}
\max_{\S_t} \quad & \EE\Bigg[ \sum_{n=1}^N R^+_{n,t}(S_{n,t},D_{n,t})\Bigg] \label{eq:single-period-problem-a}\\
\text{s.t.} \quad & S_{n,t} \;\geq\; I_{n,t}, \quad \forall n\in[N], \quad  \sum_{n=1}^N S_{n,t} \;\leq\; B.  \label{eq:single-period-problem-c}
\end{align}
\end{subequations}
Although the myopic policy serves as a heuristic in the general non-stationary case, we focus on settings where demand distributions evolve gradually (as formalized later in Assumption~\ref{asp:no_overshoot}). Under such conditions, the myopic policy effectively captures the system dynamics with minimal loss of optimality compared to the intractable global solution.


In the multi-item setting, the shared capacity constraint prevents base-stock levels from always reaching these unconstrained optimal values, creating interference among items.
To solve this constrained optimization problem, we introduce a Lagrangian multiplier $\lambda_t\ge 0$ associated with the capacity constraint \eqref{eq:single-period-problem-c}.

\begin{proposition}\label{prop:policy}
For any period $t$ with a feasible initial inventory vector, $\sum_{n=1}^N I_{n,t}\le B$, the myopic clairvoyant policy follows a modified newsvendor solution. We use the convention
\(
F_{n,t}^{-1}(q):=\inf\{s\in\mathbb R:F_{n,t}(s)\ge q\},\ q\in(0,1],
\)
and define the lower support by
\(
F_{n,t}^{-1}(0):=\inf\{s\in\mathbb R:F_{n,t}(s)>0\}.
\)
For a given Lagrange multiplier $\lambda_t\ge0$, the item-level Lagrangian maximizer is
\begin{align}
\tag{3a}
\tilde S_{n,t}(\lambda_t)
=
\begin{cases}
\displaystyle
\max\left\{I_{n,t},\; F_{n,t}^{-1}\left(\frac{m_n-\lambda_t}{M_n}\right)\right\}, & 0\le \lambda_t<m_n,\\[8pt]
I_{n,t}, & \lambda_t>m_n.
\end{cases}
\label{eq:optimal_exact}
\end{align}
When $\lambda_t=m_n$, the item-level maximizer may be nonunique; any level in
$\left[I_{n,t},\max\{I_{n,t},F^{-1}_{n,t}(0)\}\right]$ is Lagrangian-optimal. The optimal multiplier $\lambda_t^*$ is chosen so that the capacity constraint and complementary slackness hold:
\begin{align}
\tag{3b}
\lambda_t^*\ge0,\qquad
\sum_{n=1}^N S^*_{n,t}(\lambda_t^*)\le B,\qquad
\lambda_t^*\left(B-\sum_{n=1}^N S^*_{n,t}(\lambda_t^*)\right)=0,
\label{eq:lagrange_exact}
\end{align}
where $S^*_{n,t}(\lambda_t^*)$ is selected from the maximizer set above.
\end{proposition}


The solution follows the standard Lagrangian relaxation approach \citep{hadley1963analysis}. Although the cases $\lambda_t\ge m_n$ are mathematically well-defined, they induce flat or corner item-level responses and require tracking active items across assignments and forecast realizations. 
Thus we focus on the cases when multiplier stays below each unit margin and the inventory lower bound is inactive, so the exact response reduces to the newsvendor target used in our analysis.

\subsection{Demand Estimation}\label{sec:estimation}

While the true demand distributions are unknown in practice, the decision maker must estimate them in order to implement an inventory policy. To facilitate our analysis, we assume that the true demand of item $n$ in period $t$ follows a uniform distribution over the interval $\big[\mu_{n,t}-\alpha_{n,t},\, \mu_{n,t}+\alpha_{n,t}\big]$:
\[
D_{n,t} \sim \mathrm{Unif}\big[\mu_{n,t}-\alpha_{n,t},\, \mu_{n,t}+\alpha_{n,t}\big].
\]
In practice, the true parameters $(\mu_{n,t}, \alpha_{n,t})$ are unknown to the decision maker. Instead, the policy is based on an estimated demand distribution,
$
\hat{D}_{n,t} \sim \mathrm{Unif}\big[\hat\mu_{n,t}-\hat\alpha_{n,t},\, \hat\mu_{n,t}+\hat\alpha_{n,t}\big]$
constructed from historical data or auxiliary information.

Consequently, the decision maker applies the myopic single-period solution based on the estimated distribution for period $t$. The exact estimated response has the same piecewise form as Proposition~\ref{prop:policy}, with $F_{n,t}$ replaced by $\hat F_{n,t}$.   For the uniform estimated demand distribution, if the estimated Lagrange multiplier is margin-bounded (i.e., $\hat{\lambda}_t < \min_{n\in[N]} m_n$), then the estimated base-stock levels $\hat{S}_{n,t}$ are given by:
\begin{align}\label{eq:estimated}
\begin{aligned}
    & \hat S_{n,t}
= \max\Big\{
I_{n,t},\;   
\hat q_{n,t}
\Big\}, \ \hat q_{n,t}
:=\hat\mu_{n,t}+\hat\alpha_{n,t}
\left(2\frac{m_n-\hat\lambda_t}{M_n}-1\right), \quad  \forall n\in[N],\\
& \text{subject to } \hat\lambda_t \;\ge\; 0,\qquad 
\sum_{n=1}^N \hat S_{n,t} \;\le\; B,\qquad
\hat\lambda_t \Big( B - \sum_{n=1}^N \hat S_{n,t} \Big) = 0.
\end{aligned}
\end{align}

\begin{remark}
    We focus on these particular uniform distribution families to derive clear analytical results and  indeed, as shown by the computational experiments in \cite{erlebacher2000optimal}, the optimal solution obtained under the uniform distribution serves as an effective heuristic for the general case, especially at higher levels of capacity.
\end{remark}

\section{Treatment Effect Estimator and Experimental Designs}\label{section:design}

Forecasting models are often updated as new data and features become available. Because ordering decisions depend on forecasts, we compare a treatment forecasting method with a control method through the item--period outcomes induced by their ordering decisions.

Let $w\in\{0,1\}$ denote the forecasting method, where $w=0$ is the control and $w=1$ is the treatment. For item $n\in[N]$ in period $t\in[T]$, let $\hat\mu_{n,t}(w)$ be the estimated demand mean under method $w$. The two regimes in Section~\ref{sec:analysis} distinguish whether treatment affects the forecast through a systematic mean shift or through forecast-error dispersion. Focusing on misspecification in the forecast means $\{\hat\mu_{n,t}\}_{n\in[N],t\in[T]}$, we assume the demand-range parameter is correctly estimated: $\hat\alpha_{n,t}=\alpha_{n,t}$ for all $n\in[N],\,t\in[T]$.

The allocation of forecast methods is governed by a treatment assignment matrix
\[
\W=(W_{n,t})_{n\in[N],\,t\in[T]}\in\{0,1\}^{N\times T},
\]
where $W_{n,t}=1$ assigns item $n$ in period $t$ to the treatment forecast and $W_{n,t}=0$ to the control forecast.

\textbf{Global treatment effect.}
The primary metric of interest is the global treatment effect (GTE), the difference between the average outcomes when all item--period cells are exposed to treatment and when they are all exposed to control:
\[
\mathrm{GTE}
:=
\frac{1}{NT} \sum_{n=1}^N \sum_{t=1}^T
\mathbb{E}\big[ R_{n,t} \mid \W = \mathbf{1}_{N\times T} \big]
-
\frac{1}{NT} \sum_{n=1}^N \sum_{t=1}^T
\mathbb{E}\big[ R_{n,t} \mid \W = \mathbf{0}_{N\times T} \big].
\]

\textbf{Experimental design.}
The design of experiments (also known as the randomization distribution or simply the design) is a probability distribution over assignment matrices,
\[
\mathbb{P}_{\W}: \{0,1\}^{N\times T}\to[0,1],
\qquad
\sum_{\bw} \mathbb{P}_{\W}(\bw)=1,
\]
chosen by the experimenter before outcomes are observed.
We restrict attention to designs with common marginal treatment probability $\PP(W_{n,t}=1)=p, \ p\in(0,1)$ for all $n\in[N],t\in[T]$. An experiment draws one assignment matrix from $\PP_{\W}$, implements it, and estimates the causal estimand from the realized assignment and the observed outcomes.

\textbf{IPW estimator.}
Because each item--period outcome is observed under only one realization $\bw$ of $\W$, we estimate the GTE using the inverse-probability-weighted (IPW), or Horvitz--Thompson, estimator \citep{horvitz1952generalization}:
\[
\widehat{\mathrm{GTE}}(\PP_{\W},\bw)
=
\frac{1}{NT}
\sum_{n=1}^N \sum_{t=1}^T
\left(
\frac{w_{n,t} R_{n,t}}{p}
-
\frac{(1-w_{n,t}) R_{n,t}}{1-p}
\right).
\]
Under the stable unit treatment value assumption (SUTVA) this estimator is unbiased for the GTE. With temporal carryover or cross-item interference, however, it is generally biased. We define its bias as
\[
\text{Bias}:= \EE[\widehat{\mathrm{GTE}}(\PP_{\W},\bw)]-\mathrm{GTE},
\]
where the expectation is over both the assignment mechanism and exogenous demand or operational randomness.
We characterize its sign and magnitude as a function of the design.

We compare three designs, shown in Figure~\ref{fig:designs}, that have the same marginal treatment probability $p$ but different dependence across items and periods.

\textbf{Switchback (SW).}
The switchback design assigns all items to the same method within a period and rerandomizes across periods. Let $Z_t\overset{i.i.d.}{\sim}\mathrm{Bernoulli}(p)$ and set
\[
W_{n,t}=Z_t,\qquad n\in[N],\ t\in[T].
\]
This period-level randomization is natural when capacity or substitution creates strong within-period interference, but it remains sensitive to temporal carryover \citep{bojinov2023design}.

\begin{figure}[!ht]
\centering

\begin{minipage}[t]{0.32\columnwidth}
\centering
\resizebox{\linewidth}{!}{%
$
\begin{array}{c}
\text{Time } t\\[-2pt]
\left(
\begin{array}{c|cccccccc}
\text{Item 1} & 1&1&0&1&0&0&1&1\\
\text{Item 2} & 1&1&0&1&0&0&1&1\\
\text{Item 3} & 1&1&0&1&0&0&1&1\\
\text{Item 4} & 1&1&0&1&0&0&1&1\\
\text{Item 5} & 1&1&0&1&0&0&1&1\\
\text{Item 6} & 1&1&0&1&0&0&1&1
\end{array}
\right)
\end{array}
$
}
\subcaption{SW experiment}
\label{fig:SW}
\end{minipage}\hfill
\begin{minipage}[t]{0.32\columnwidth}
\centering
\resizebox{\linewidth}{!}{%
$
\begin{array}{c}
\text{Time } t\\[-2pt]
\left(
\begin{array}{c|cccccccc}
\text{Item 1} & 1&1&1&1&1&1&1&1\\
\text{Item 2} & 1&1&1&1&1&1&1&1\\
\text{Item 3} & 0&0&0&0&0&0&0&0\\
\text{Item 4} & 0&0&0&0&0&0&0&0\\
\text{Item 5} & 1&1&1&1&1&1&1&1\\
\text{Item 6} & 0&0&0&0&0&0&0&0
\end{array}
\right)
\end{array}
$
}
\subcaption{IR experiment}
\label{fig:IR}
\end{minipage}\hfill
\begin{minipage}[t]{0.32\columnwidth}
\centering
\resizebox{\linewidth}{!}{%
$
\begin{array}{c}
\text{Time } t\\[-2pt]
\left(
\begin{array}{c|cccccccc}
\text{Item 1} & 1&0&1&1&0&1&0&0\\
\text{Item 2} & 0&1&1&0&0&1&1&0\\
\text{Item 3} & 1&1&0&0&1&0&1&1\\
\text{Item 4} & 0&0&1&1&1&0&0&1\\
\text{Item 5} & 1&0&0&1&0&1&1&0\\
\text{Item 6} & 0&1&1&0&1&1&0&0
\end{array}
\right)
\end{array}
$
}
\subcaption{PR experiment}
\label{fig:PR}
\end{minipage}

\caption{Examples of experimental designs. Each column corresponds to one time period $t$. In SW every column is constant across items; in IR every row is constant across periods; in PR every cell is drawn independently.}
\label{fig:designs}
\end{figure}

\textbf{Item-level Randomization (IR).}
The item-level design fixes each item's assignment over the entire horizon. Let $Z_n\overset{i.i.d.}{\sim}\mathrm{Bernoulli}(p)$ and set
\[
W_{n,t}=Z_n,\qquad n\in[N],\ t\in[T].
\]
This preserves a coherent treatment path for each item, but it can be biased when treated and control items interact through shared inventory capacity or substitution \citep{blake2014marketplace,johari2022experimental}.

\textbf{Pairwise Randomization (PR).} The pairwise design independently randomizes every item--period cell:
\[
W_{n,t}\overset{i.i.d.}{\sim}\mathrm{Bernoulli}(p),
\qquad n\in[N],\ t\in[T].
\]
This design removes dependence across both items and periods by construction, but the resulting mixed assignments may aggravate interference and carryover in inventory systems.

\section{Bias Analysis under Different Interventions}\label{sec:analysis}

This section studies A/B tests that compare a control demand forecast with a treatment demand forecast in a multi-item, multi-period inventory system with shared capacity. Each assignment determines which forecast is used for each item-period, which then affects replenishment decisions and rewards. We compare the true global treatment effect with the IPW estimates produced by switchback, item-level, and pairwise randomization designs, and characterize the resulting design-specific biases. We focus on two forecasting improvements: reducing downward bias in forecast means and reducing forecast-error dispersion.

\begin{itemize}
\item \textbf{Scenario 1.} The treatment reduces a common downward bias in forecast means, as in lost-sales systems where censored sales lead naive estimators to underestimate demand \citep{nahmias1994demand}. This captures censoring-correction methods, such as Kaplan--Meier-type estimators \citep{huh2009adaptive}. We characterize the GTE and the bias of the IPW estimator under SW, IR, and PR designs.

\item \textbf{Scenario 2.} The treatment preserves the mean but reduces forecast-error dispersion, capturing forecasting upgrades that improve precision without changing average predictions. We derive large-$N$ limits for the GTE and the associated design-specific biases.
\end{itemize}

Together, the results show that experimental bias depends sharply on both the design and the type of forecasting improvement. Proofs are given in Appendices~\ref{appendix:scenario1-proofs} and~\ref{appendix:scenario2-proofs}.










\begin{table}[t]
\centering
\small
\renewcommand{\arraystretch}{1.2} 
\begin{tabular*}{\textwidth}{@{\extracolsep{\fill}}l c c c c}
\toprule
\textbf{Scenario} & \textbf{True GTE} & \textbf{SW Bias} & \textbf{IR Bias} & \textbf{PR Bias} \\
\midrule
Reduce downward bias & Positive & Negative & Positive & Upper Bounded by IR \\
\hline
Reduce forecast-error dispersion & Asymp. Positive & Asymp. Positive & Asymp. Unbiased & Asymp. Positive \\
\bottomrule
\end{tabular*}
\caption{Summary of the theoretical signs for the Global Treatment Effect (GTE) and the estimation bias of different randomization designs.}
\label{tab:summary_of_results}
\end{table}

\subsection{Reducing Downward Bias in Forecast Means}\label{section:scenario1}

We first study how demand-estimation bias affects A/B tests. In lost-sales systems, demand is often underestimated because sales are censored: once inventory is depleted, excess demand is unobserved. We therefore ask whether a more accurate forecasting method can mitigate this downward bias by accounting for censoring, and whether different experimental designs accurately recover the resulting treatment effect.

We consider a regime in which both control and treatment forecasts underestimate the true mean demand, but the treatment is less biased. To isolate mean bias, we assume zero forecast dispersion, equivalently that the forecast mean is deterministic conditional on $w$, for all $w\in\{0,1\}$, $n\in[N]$, and $t\in[T]$. Forecast-error dispersion is analyzed separately in Section~\ref{section:scenario2}.

\vspace{5pt}

\textit{\emph{Scenario 1:} For each item $n \in [N]$, period $t \in [T]$, there exists a forecast error parameter $\Delta$ such that the estimated demand mean is given by:
\[
\hat \mu_{n,t}^{(\Delta)} = \mu_{n,t} + \Delta \alpha_{n,t}.
\]
We consider the case where the treatment and control satisfy
\[
\hat \mu_{n,t} (0) = \hat \mu_{n,t}^{(\Delta_0)},\ \hat \mu_{n,t} (1) = \hat \mu_{n,t}^{(\Delta_1)}, \ \Delta_0 < \Delta_1 \le 0.
\]}

\vspace{5pt}

Let $\lambda_t^*$ denote the myopic KKT multiplier in period $t$ under the true demand distribution.


\begin{assumption}[Nonnegative estimated support and no inventory overshoot]
\label{asp:no_overshoot}
For every experiment considered in the analysis, define the affine newsvendor target
\begin{align*}
\bar S_{n,t}(\W_t)
:=\hat\mu_{n,t}(W_{n,t})
+\hat\alpha_{n,t}
\left(2\frac{m_n-\hat\lambda_t(\W_t)}{M_n}-1\right).
\end{align*}
We assume the following two conditions hold for every realization of the treatment assignment $\boldsymbol{w} \in \{0,1\}^{N\times T}$ and forecast estimates considered in the theoretical analysis:
\begin{enumerate}
    \item[(i)]The estimated lower support is nonnegative:
    \[
    \hat\mu_{n,t}(w_{n,t})-\hat\alpha_{n,t} \ge0,
    \qquad \forall n\in[N],\ t\in[T].
    \]
    \item[(ii)] For any $n\in[N]$ and $t\in[T-1]$,
    \[
    \bar S_{n,t}(\boldsymbol{w}_t)-\bar S_{n,t+1}(\boldsymbol{w}_{t+1})\le \underline D_{n,t},
    \]
    where $\underline D_{n,t}=\mu_{n,t}-\alpha_{n,t}$ is the lower support of the true demand distribution $D_{n,t}$.
\end{enumerate}
\end{assumption}

In the uniform demand setting, Assumption~\ref{asp:no_overshoot} holds naturally when demand does not exhibit abrupt fluctuations across adjacent periods, and when the lower support of the demand distribution, $\underline D_{n,t}=\mu_{n,t}-\alpha_{n,t}$, is sufficiently large relative to the magnitude of forecast-induced changes in the target base-stock levels, which are of order $O(\alpha_{n,t})$.

Assumption~\ref{asp:no_overshoot} rules out overshooting order-up-to targets across periods: part~(i) initializes the induction with $I_{n,1}=0\le \bar S_{n,1}$, and part~(ii) gives the induction step. Together with Assumption~\ref{asp:margin_bounded_multiplier}, it implies that the inventory lower bound in~\eqref{eq:estimated} is inactive and $\hat S_{n,t}=\bar S_{n,t}$ for all $n,t$; see Lemma~\ref{lem:s1_affine_response} in Appendix~\ref{appendix:scenario1-proofs}. Thus, the current decision depends only on the current assignment $\W_t$, not on past assignments $\W_{N\times(t-1)}$, isolating the direct effect of the current forecast. We therefore write the base-stock levels and myopic multiplier as $\hat S_{n,t}(\W_t)$ and $\hat\lambda_t(\W_t)$. In the SW experiment, when all items share forecast-error parameter $\Delta$ at time $t$, we write $\hat S_{n,t}^{(\Delta)}$ and $\hat\lambda_t^{(\Delta)}$. Past assignments still affect current profit $R_{n,t}$ through the initial inventory $I_{n,t}$.

\begin{assumption}[Margin-bounded true-demand multiplier]
\label{asp:margin_bounded_multiplier}
For every period, the true-demand multiplier satisfies
\begin{equation}
\tag{MB}\label{eq:margin_bounded_multiplier}
\lambda_t^* < \min_{n\in[N]} m_n,
\qquad \forall t\in[T].
\end{equation}
\end{assumption}
Assumption~\ref{asp:margin_bounded_multiplier} rules out the knife-edge case $\lambda_t=m_n$ and the scarce-capacity case $\lambda_t>m_n$ in the true-demand system. 
In Scenario~1, the downward-bias condition $\Delta_0<\Delta_1\le0$ means the estimated systems have weakly lower aggregate demand pressure than the true-demand system. Thus, for every assignment vector $\boldsymbol w\in\{0,1\}^N$, the KKT multiplier satisfies
\[
\hat\lambda_t(\boldsymbol w)\le \lambda_t^*, \quad \forall t \in [T].
\]
This ensures that the margin-bounded regime carries over from the true-demand system to all estimated systems. Together with Assumption~\ref{asp:no_overshoot}, these conditions imply that the inventory lower bound is inactive, so the implemented base-stock levels coincide with their affine targets $\bar S_{n,t}$. The next lemma shows that, when the forecast bias is not too large, the solution of the estimated-demand system is either weakly slack or strictly interior, thereby excluding the partially binding case in which the multiplier is positive but at least one item's base-stock level hits the lower support of its demand distribution.

\begin{lemma}[Existence of a common structural interval]
\label{lem:Delta0} In Scenario 1,  
under Assumptions~\ref{asp:no_overshoot} and ~\ref{asp:margin_bounded_multiplier}, there exist period-specific thresholds $\widetilde\Delta_t<0$ and a common lower bound $\underline{\Delta}:=\max_t\widetilde\Delta_t<0$ such that, for every $t\in[T]$ and every $\Delta\in(\underline{\Delta},0]$, the solution~\eqref{eq:estimated} is either
\begin{itemize}
\item weakly slack: $\hat \lambda_t^{(\Delta)}=0$ and $\sum_n \hat S_{n,t}^{(\Delta)} \le B$, or
\item strictly interior: $0<\hat \lambda_t^{(\Delta)}<\min_{n\in[N]} m_n$, and $\hat S_{n,t}^{(\Delta)} >  \mu_{n,t} - \alpha_{n,t}$ for all $n$.
\end{itemize}
\end{lemma}

Our first result characterizes the monotonic effect of reducing the forecast bias on expected profits. In particular, it shows that, when the forecast error parameter lies within the interval $(\underline{\Delta},0]$, better demand estimates always lead to higher expected profits.

\begin{proposition}    [Sign of the GTE]
\label{thm:S1_gte} In Scenario 1, 
under Assumptions~\ref{asp:no_overshoot} and ~\ref{asp:margin_bounded_multiplier}, the global
treatment effect is
\[
GTE 
= \frac{1}{NT}\sum_{n=1}^N\sum_{t=1}^T 
\Bigl( \EE R_{n,t}^+(\hat S_{n,t}(\boldsymbol{1}),D_{n,t})
      -\EE R_{n,t}^+(\hat S_{n,t}(\boldsymbol{0}),D_{n,t}) \Bigr).
\]
Let $\underline{\Delta}$ be the bound from 
Lemma~\ref{lem:Delta0}. Then $GTE$ is nonnegative whenever
\[
\underline{\Delta} \;<\; \Delta_0 < \Delta_1 \le 0.
\]
\end{proposition}

Proposition~\ref{thm:S1_gte} shows that reducing downward bias $(\Delta_1>\Delta_0)$ increases expected profit on $(\underline{\Delta},0]$. This sign result does not require all order-up-to levels to increase: capacity reallocation may raise some items' targets and lower others. It relies instead on the system remaining in the same structural case of Lemma~\ref{lem:Delta0}, so the optimal solution varies continuously with $\Delta$.



We next analyze design-specific IPW bias. In a switchback experiment, inventory carryover links current rewards to previous treatment assignments. The following theorem expresses the SW bias as a leftover-inventory term and gives a sufficient condition under which this bias is non-positive, so that SW conservatively estimates the GTE.

\begin{theorem}[Bias in SW experiment]
\label{thm:S1_sw}
In Scenario~1, under Assumptions~\ref{asp:no_overshoot} and ~\ref{asp:margin_bounded_multiplier}, the bias of the IPW estimator in the
switchback experiment is
\[
{Bias}^{SW}
= -\frac{1}{NT}\sum_{n=1}^N\sum_{t=1}^{T-1}
c_n\Bigl( \EE(\hat S_{n,t}(\boldsymbol{1})-D_{n,t})^+
          -\EE(\hat S_{n,t}(\boldsymbol{0})-D_{n,t})^+ \Bigr).
\]
Let $\{\widetilde\Delta_t\}_{t=1}^T$ be the period-specific lower bounds from
Lemma~\ref{lem:Delta0}, and let $\underline{\Delta}:=\max_t\widetilde\Delta_t$ be the common lower bound. Suppose, in addition, the period-specific bounds satisfy
\begin{align}\label{asp:S1-sw-lowerbound}
    \theta_t \widetilde\Delta_t + \kappa_t\leq0, \ \forall t \in [T],
\end{align}
where $\theta_t, \kappa_t$ are defined in \eqref{eq:at-bt-appendix}. Then for all
\(\underline{\Delta}< \Delta_0 < \Delta_1 \le0 \), the switchback bias is nonpositive:
\[
{Bias}^{SW}\;\le\;0.
\]
\end{theorem}

The bias expression in Theorem~\ref{thm:S1_sw} shows that the SW bias is governed by how forecast improvements affect expected leftover inventory. If reducing downward bias mainly raises order-up-to levels, treatment generates more leftover inventory, which can carry into later control periods and make the IPW contrast conservative. If capacity is tight, however, forecast improvements may mostly reallocate a fixed inventory budget across items, so leftover can decrease and the bias need not be negative. The condition $\theta_t\widetilde\Delta_t+\kappa_t\le0$ rules out this reversal by ensuring that the weighted leftover term is nondecreasing on $(\underline{\Delta},0]$, yielding ${Bias}^{SW}\le0$.

In contrast, IR fixes each item’s assignment over time, eliminating temporal switching but creating persistent cross-sectional heterogeneity. Under the same demand-bias scenario, the next theorem shows that IR has nonnegative bias.

\begin{theorem}[Bias in IR experiment]\label{thm:S1_IR}
    In Scenario~1, under Assumptions~\ref{asp:no_overshoot} and ~\ref{asp:margin_bounded_multiplier}, in the IR experiment, the bias of the IPW estimator is 
\begin{align*}
Bias^{IR}= & \frac{1}{NT}\sum_{n=1}^N \sum_{t=1}^T\Big(\EE[ R_{n,t}^+(\hat S_{n,t}(\W_{t}),D_{n,t}) \mid W_{n,t} = 1] - \EE R_{n,t}^+(\hat S_{n,t}(\boldsymbol{1}),D_{n,t}) \Big) \\
   & - \frac{1}{NT}\sum_{n=1}^N \sum_{t=1}^T\Big(\EE[ R_{n,t}^+(\hat S_{n,t}(\W_{t}),D_{n,t}) \mid W_{n,t} = 0] - \EE R_{n,t}^+(\hat S_{n,t}(\boldsymbol{0}),D_{n,t}) \Big),
\end{align*}
    which is non-negative: 
$
Bias^{IR} \geq 0.
$
\end{theorem}

The nonnegative bias of IR is driven by the capacity constraint. 
With binding capacity, treated items’ higher targets crowd out control items relative to the global-control counterfactual, while treated items face weaker competition than under global treatment. 
Both inflate the treated–control contrast and therefore produce a systematically upward-biased estimator.

The previous results show that SW and IR sit on opposite sides of the truth.
Pairwise Randomization (PR) blends the two dependence structures that drive the biases in SW and IR. 
Like SW, it generates temporal dependence in inventory carryover, inducing a downward force on the estimator; like IR, it creates cross-sectional competition under a binding capacity constraint, inducing an upward force. Because these forces operate in opposite directions, the resulting bias is upper-bounded by the bias under IR.

\begin{theorem}[Bias in PR experiment]\label{thm:S1_PR}
    In Scenario~1, under Assumptions~\ref{asp:no_overshoot} and ~\ref{asp:margin_bounded_multiplier}, the bias of the IPW estimator in the pairwise randomized experiment is
    \begin{align*}
     Bias^{PR} & =
            \frac{1}{NT}\sum_{n=1}^N \sum_{t=1}^T\Big(\EE[ R_{n,t}^+(\hat S_{n,t}(\W_{t}),D_{n,t}) \mid W_{n,t} = 1] - \EE R_{n,t}^+(\hat S_{n,t}(\boldsymbol{1}),D_{n,t}) \Big)  \\
       & - \frac{1}{NT}\sum_{n=1}^N \sum_{t=1}^T\Big(\EE[ R_{n,t}^+(\hat S_{n,t}(\W_{t}),D_{n,t}) \mid W_{n,t} = 0] - \EE R_{n,t}^+(\hat S_{n,t}(\boldsymbol{0}),D_{n,t}) \Big)  \\
   & -  \frac{1}{NT} \sum_{n=1}^N \sum_{t=1}^{T-1} c_n \Big(\EE[(\hat S_{n,t}(\W_{t}) - D_{n,t})^+ \mid W_{n,t} = 1]  - \EE[(\hat S_{n,t}(\W_{t}) - D_{n,t})^+ \mid W_{n,t} = 0] \Big), 
\end{align*}
which is upper bounded by the bias in the IR experiment, i.e.,
$
Bias^{PR} \leqslant Bias^{IR}.
$
\end{theorem}


PR combines cross-sectional competition (IR-type upward force) with inventory carryover (SW-type downward force), yielding $Bias^{PR}\le Bias^{IR}$ in Theorem~\ref{thm:S1_PR}. Notice that $Bias^{PR}$ is not necessarily lower-bounded by $Bias^{SW}$,  {as PR also amplifies inventory level variations because of larger discrepancies in concurrent base-stock levels (see ~\eqref{eq:basestock_treat_vs_GT}~\eqref{eq:basestock_control_vs_GC} in Appendix~\ref{appendix:scenario1-proofs}):
\[
\PP[\hat S_{n,t}(\boldsymbol{\W_t}) \ge \hat S_{n,t}(\boldsymbol{1}) \mid W_{n,t}=1] = 1,\ \PP[\hat S_{n,t}(\boldsymbol{0}) \ge \hat S_{n,t}(\boldsymbol{\W_t}) \mid W_{n,t}=0] = 1
\]}
and thus produces a larger downward-biased effect.

\subsection{Reducing Forecast-Error Dispersion with Unchanged Mean}\label{section:scenario2}
\label{section:scenario2}

We next study A/B tests for forecast improvements that reduce error dispersion without changing the mean forecast. This regime captures upgrades that make demand predictions more stable while preserving the average prediction, a common pattern for modern machine-learning forecasters.

\vspace{5pt}

\textit{\emph{Scenario 2:} For each item $n\in[N]$, period $t\in[T]$, and group $w\in\{0,1\}$,
\[
\hat\mu_{n,t}(w)=\mu_{n,t}+\epsilon_{n,t}(w).
\]
We consider the case where treatment and control satisfy that, $\forall n\in[N], t\in[T], w\in\{0,1\}$,
\[
\EE[\epsilon_{n,t}(w)]=0,
\qquad
\epsilon_{n,t}(1)\le_{\mathrm{cx}}\epsilon_{n,t}(0).
\]}

\vspace{5pt}

The convex-order condition formalizes a dispersion-reduction improvement: it implies that every convex loss of the forecast error is weakly smaller under treatment. In particular, whenever the variances exist,  $\Var(\epsilon_{n,t}(1))\le \Var(\epsilon_{n,t}(0))$.


Modern large-scale inventory systems often involve thousands or millions of items sharing a joint warehouse capacity. This motivates a mean--field regime in which $N$ and $B^{(N)}$ grow proportionally and cross-sectional averages converge to deterministic limits. Let $\hat\lambda_t^{(N)}(\W_t)$ and $\hat S_{n,t}^{(N)}(\W_t)$ denote the predictive KKT multiplier and base-stock level in the $N$-item system. 
Unlike Scenario~1, the margin-bounded property cannot be inherited from a deterministic downward-bias ordering, because the forecasts are random. The next assumption is therefore the Scenario~2 analogue of Assumption~\ref{asp:margin_bounded_multiplier}: it keeps the entire bounded-error class inside the same affine-response region.

\begin{assumption}\label{asp:s2_margin}
There exists a deterministic radius $\eta>0$ such that $|\epsilon_{n,t}(w)|\le\eta$ a.s.\ for all $n,t,w$, and, uniformly over every $N$, period $t$, assignment $\W_t$, and forecast-error realization in this $\eta$-class, the predictive system satisfies
\[
0\le\hat\lambda_t^{(N)}(\W_t)<m_{\min}^{(N)}:=\min_{n\in[N]}m_n  \ a.s..
\]
\end{assumption}

Under Assumption~\ref{asp:no_overshoot} and Assumption~\ref{asp:s2_margin}, the inventory lower bound in~\eqref{eq:estimated} is inactive and the predictive base-stock level admits the affine representation
\[
\hat S_{n,t}^{(N)}(\W_t)
=
\bar S_{n,t}^{(N)}(\W_t)
=
\hat\mu_{n,t}(W_{n,t})
+\alpha_{n,t}\Bigl(2(u_n-v_n\hat\lambda_t^{(N)}(\W_t))-1\Bigr),
\]
where $u_n$ and $v_n$ are defined in Section~\ref{sec:dynamics}. Appendix~\ref{app:s2-margin-verification} gives conditions on $\eta$ that imply Assumption~\ref{asp:s2_margin}.

The remaining assumptions are used for the mean-field limit: independence gives a law of large numbers, bounded moments give uniform integrability, and the two limit assumptions ensure that the aggregate capacity slack has a unique deterministic crossing.
\begin{assumption}[Independence]\label{asp:independence}
For all $n\in[N],t\in[T]$ and $w\in\{0,1\}$:
\begin{enumerate}
    \item The pairs of forecast errors $\{(\epsilon_{n,t}(0),\epsilon_{n,t}(1))\}_{n}$ are independent across items $n$.
    \item The assignment process $\{W_{n,t}\}_{n,t}$ is independent of
the demand process $\{D_{n,t}\}_{n,t}$ and the forecasting signals
$\{\epsilon_{n,t}(w)\}_{n,t,w}$. Moreover, for all $n,t$ and $w\in\{0,1\}$,
$\epsilon_{n,t}(w)\perp\!\!\!\perp (D_{n,t}-\mu_{n,t})$.
\end{enumerate}
\end{assumption}
Item-level independence ensures that cross-sectional averages converge
by the law of large numbers, which is essential for the mean-field limit.
The independence between forecast errors and demands ensures that forecast uncertainty impacts profits solely through the stocking decision rather than through spurious correlations, which allows us to use convex-order comparisons in the proofs for GTE and bias.

\begin{assumption}[Uniform moment and parameter bounds]\label{asp:moment_bounds}
\begin{enumerate}
    \item $\sup_{n,t}(|\mu_{n,t}|+|\alpha_{n,t}|)<\infty$;
\item The sequences $\{b_n\}_n$, $\{c_n\}_n$, and $\{h_n\}_n$ are uniformly bounded, $b_n > c_n\ge 0$ and $h_n\ge 0$ for all $n$, and $\inf_n m_n>0$.

\end{enumerate}
\end{assumption}
These bounds rule out extreme items with unbounded demand or costs and, together with the bounded-error part of Assumption~\ref{asp:s2_margin}, guarantee the integrability of revenues and leftover inventories in the proofs. 

\begin{assumption}[Parameter limits]\label{asp:para_bounded_limit}
There exist finite constants $\beta_0,\mu_0,\alpha_0,u_0,v_0$ such that, 
\begin{align*}
 \text{for each fixed $t\in[T]$,}   & \lim_{N\to\infty}\frac{B^{(N)}}{N} = \beta_0, \lim_{N\to\infty}
\frac1N\sum_{n=1}^N \mu_{n,t}=\mu_0,
\lim_{N\to\infty}\frac1N\sum_{n=1}^N \alpha_{n,t}=\alpha_0,\\
& \lim_{N\to\infty}\frac1N\sum_{n=1}^N \alpha_{n,t}u_n=\alpha_0 u_0, 
\lim_{N\to\infty}\frac1N\sum_{n=1}^N \alpha_{n,t}v_n=\alpha_0 v_0.
\end{align*}
\end{assumption}

\begin{assumption}[Mean-field limit of base-stock responses]\label{asp:kt_limit}
Let
\[
m_*:=\inf_{n\ge1}m_n>0.
\]
For each fixed $\lambda\in[0,m_*)$, the limit
\begin{align*}
k_t(\lambda)
:=\lim_{N\to\infty}\frac1N\sum_{n=1}^N 
\alpha_{n,t}\Bigl(2(u_n-v_n\lambda)-1\Bigr)
\end{align*}
exists and is finite.
\end{assumption}

Assumptions~\ref{asp:para_bounded_limit}--\ref{asp:kt_limit} ensure that the aggregate capacity slack admits a deterministic mean-field limit as $N\to\infty$, so that the KKT multiplier converges to a fixed point. These conditions hold, for example, if the item parameters $\{(b_n,c_n,h_n,\mu_{n,t},\alpha_{n,t})\}_{n=1}^N$ are i.i.d. samples from a bounded distribution with margins bounded away from zero.

\begin{assumption}[Mean-field KKT multiplier]\label{asp:unique_zero}
Let
\[
\psi_t(\lambda)=\mu_0+k_t(\lambda)-\beta_0,\qquad \lambda\in[0,m_*).
\]
 Define
\[
\lambda_t^*
:=
\inf\{\lambda\in[0,m_*):\psi_t(\lambda)\le0\}.
\]
We assume that the set $\{\lambda\in[0,m_*):\psi_t(\lambda)\le0\}$ is nonempty and that
$\lambda_t^*<m_*$. Moreover, $\psi_t$ crosses the KKT threshold uniquely in the following sense:
\begin{itemize}
    \item if $\psi_t(0)>0$, then $\psi_t(\lambda)>0$ for all $\lambda<\lambda_t^*$ and
    $\psi_t(\lambda)<0$ for all $\lambda>\lambda_t^*$ in $[0,m_*)$;
    \item if $\psi_t(0)\le0$, then $\lambda_t^*=0$ and $\psi_t(\lambda)<0$ for every
    $\lambda\in(0,m_*)$.
\end{itemize}
\end{assumption}


The formulation above includes both the binding case $\psi_t(0)>0$ and the slack case $\psi_t(0)\le0$, in which the limiting multiplier is $\lambda_t^*=0$. The following lemma shows that, in the mean--field limit, the random KKT multiplier becomes deterministic.

\begin{lemma}[Mean-field limit of the Lagrange multiplier]\label{lem:lambda_limit}
In Scenario 2, under Assumption~\ref{asp:no_overshoot} and  Assumptions~\ref{asp:s2_margin}--\ref{asp:unique_zero}, for each fixed $t\in[T]$,
\begin{align*}
\hat\lambda_t^{(N)}(\mathbf W_t)\xrightarrow[N\to\infty]{a.s.}\lambda_t^* \quad \text{ and } \quad
\EE\big|\hat\lambda_t^{(N)}(\mathbf W_t)-\lambda_t^*\big|\xrightarrow[N\to\infty]{}0.
\end{align*}
Consequently,
\begin{align*}
\hat S_{n,t}^{(N)}(\mathbf W_t)
\;\xrightarrow[N\to\infty]{a.s.}\;
X_{n,t}(W_{n,t})
:=\hat\mu_{n,t}(W_{n,t})
+\alpha_{n,t}\Bigl(2(u_n-v_n\lambda_t^*)-1\Bigr).
\end{align*}
The same convergence statements hold when $\mathbf W_t$ is replaced by any deterministic assignment vector independent of the forecast errors, including the global assignments $\boldsymbol{1}$ and $\boldsymbol{0}$.
\end{lemma}

Lemma~\ref{lem:lambda_limit} removes the main finite-system complication: in the mean-field limit, treatment assignments and forecast noise affect capacity only through a deterministic multiplier $\lambda_t^*$. The remaining treatment-control difference is therefore the idiosyncratic forecast-error dispersion, ordered by convex order.

We now present the main results on the GTE and the asymptotic bias of different
experimental designs.

\begin{proposition}[Asymptotic non-negative GTE]\label{thm:S2_gte}
In Scenario 2, under Assumption~\ref{asp:no_overshoot}  and Assumptions~\ref{asp:s2_margin}--\ref{asp:unique_zero},
\[
\liminf_{N\to\infty} \mathrm{GTE}^{(N)} \ge 0.
\]
\end{proposition}

In Scenario 2, the treatment reduces the dispersion of the mean forecast error while keeping the forecast centered at the true mean. 
Because $s\mapsto (s-D)^+$ is convex, the convex-order improvement $\epsilon(1)\le_{\mathrm{cx}}\epsilon(0)$ lowers expected leftover $\EE(s-D)^+$ and thus weakly increases expected profit in the mean-field limit. This is an asymptotic statement: before passing to the limit, the finite-$N$ base-stock level depends on the empirical multiplier and is capacity-coupled with the forecast errors, so the convex-order comparison is not asserted at finite $N$.

\begin{theorem}[Asymptotic non-negative bias of SW]\label{thm:S2_sw}
In Scenario 2,  under Assumption~\ref{asp:no_overshoot}  and Assumptions~\ref{asp:s2_margin}--\ref{asp:unique_zero}, 
\[
\liminf_{N\to\infty} \mathrm{Bias}^{SW,(N)}\ge 0.
\]
\end{theorem}

In Scenario~2, SW is upward biased through an intertemporal leftover-carryover channel. Economically, the switchback estimator compares treatment and control periods without fully accounting for the inventory carried over from the previous period. In the mean-field limit, the convex-order improvement makes the treatment base-stock level less dispersed and therefore lowers expected leftover inventory relative to control, thereby making the carryover contribution to the bias positive.

\begin{theorem}[Asymptotic unbiasedness of IR]\label{thm:S2_ir}
In Scenario 2, under Assumption~\ref{asp:no_overshoot}  and Assumptions~\ref{asp:s2_margin}--\ref{asp:unique_zero}, 
\[
\lim_{N\to\infty} \mathrm{Bias}^{IR,(N)}=0.
\]
\end{theorem}

In item-level randomization, each item is permanently assigned to either treatment or
control, so there is no temporal switching at the item level. In the mean-field limit, the
capacity multiplier $\lambda_t^*$ is shared by both groups and becomes deterministic. Therefore, the
environment faced by the treatment (/control) group in IR coincides asymptotically
with that in the corresponding global-treatment (/global-control) regime. Thus, the
naive IPW estimator compares two groups that differ only in forecast-error dispersion, without
additional spillover distortions, and the IR design becomes asymptotically unbiased.

\begin{theorem}[Asymptotic non-negative bias of PR]\label{thm:S2_pr}
In Scenario 2, under Assumption~\ref{asp:no_overshoot} and Assumptions~\ref{asp:s2_margin}--\ref{asp:unique_zero}, 
\[
\mathrm{Bias}^{PR,(N)}-\mathrm{Bias}^{SW,(N)}\to 0.
\]
Consequently,
\[
\liminf_{N\to\infty}\mathrm{Bias}^{PR,(N)}
=\liminf_{N\to\infty}\mathrm{Bias}^{SW,(N)}\ge 0.
\]
\end{theorem}



PR shares the same mean-field multiplier and inherits the same asymptotic intertemporal leftover-carryover term as SW, while IR remains unbiased.  The results in Scenario~2 stand in contrast to Scenario~1 and highlight that the direction of bias is sensitive to whether the treatment improves the mean or the dispersion of forecast errors.

\section{Numerical Experiments}\label{sec:numerical}

This section complements the theoretical analysis in Sections~\ref{section:scenario1} and~\ref{section:scenario2} with two sets of simulations. First, Section~\ref{subsec:theory_aligned_sim} reports controlled stochastic simulations that exactly implement the uniform-demand model used in the theory. These simulations are designed to verify the signs and mechanisms in Table~\ref{tab:summary_of_results}. Second, Section~\ref{subsec:trace_driven_sim} reports trace-driven experiments on FreshRetailNet-50K. These  experiments use recovered real demand traces, point forecasts, and a stockout-substitution module; they show that the same mechanisms persist in a more realistic retail
environment. Detailed simulation parameters are reported in Appendix~\ref{app:implementation_details}.

\subsection{Synthetic Experiment under the Uniform-Demand Model}\label{subsec:theory_aligned_sim}

We first run simulations that match the stochastic model in Section~\ref{sec:estimation}. For each item $n$ and period $t$, true demand is generated as
\[
D_{n,t}=\mu_n+\alpha_n U_{n,t},\qquad U_{n,t}\sim \mathrm{Unif}[-1,1],\qquad \alpha_n>0.
\]
The inventory policy is the myopic base-stock rule computed from the forecasted uniform distribution $(\hat\mu_{n,t},\hat\alpha_{n,t})$ and the common capacity constraint. We use the three assignment mechanisms as in Section~3: SW, IR and PR, all with marginal treatment probability $p=1/2$. For each design, we estimate the average GTE contrast over 300 independent randomizations and compare it with the Monte Carlo global-treatment/global-control benchmark. 

\begin{figure}[t]
\centering
\begin{subfigure}{0.32\textwidth}
\centering
\includegraphics[width=\linewidth]{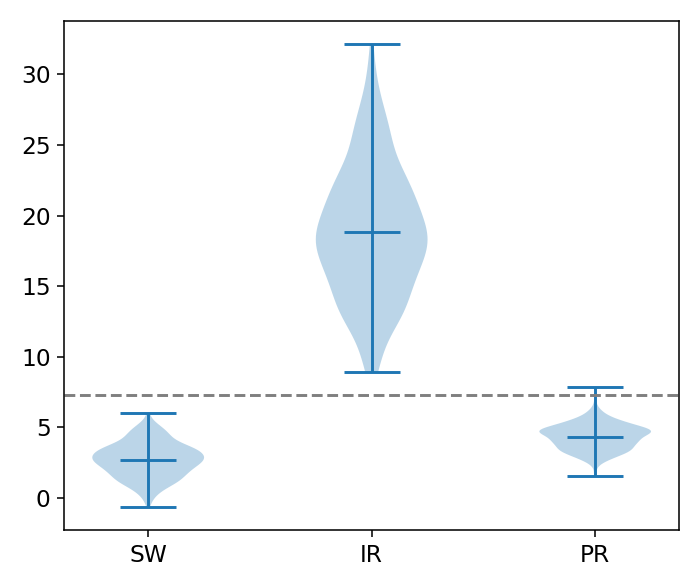}
\caption{Tight capacity}
\end{subfigure}\hfill
\begin{subfigure}{0.32\textwidth}
\centering
\includegraphics[width=\linewidth]{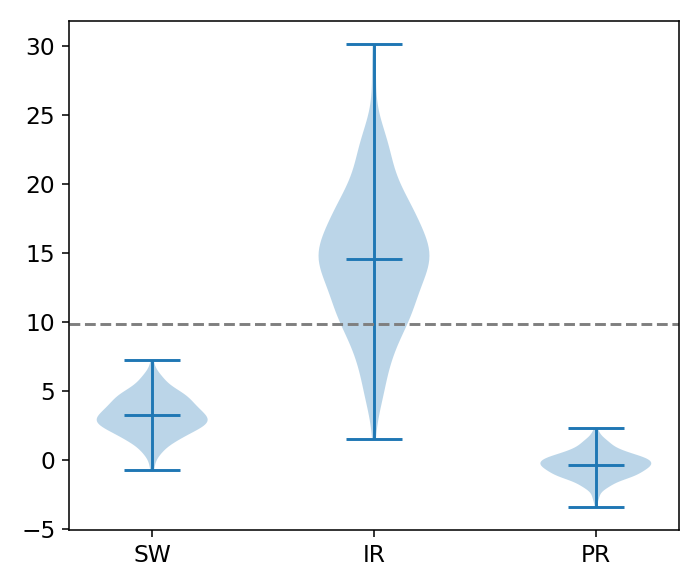}
\caption{Medium capacity}
\end{subfigure}\hfill
\begin{subfigure}{0.32\textwidth}
\centering
\includegraphics[width=\linewidth]{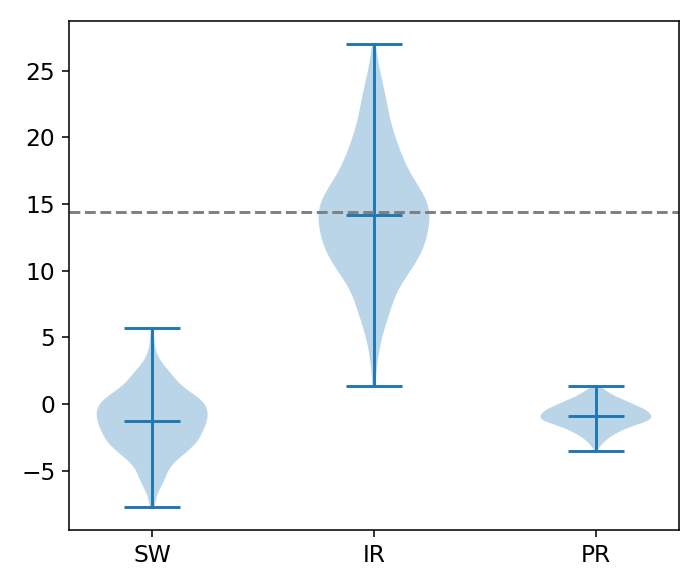}
\caption{Loose capacity}
\end{subfigure}
\caption{Simulation for Scenario~1. Treatment reduces downward mean bias. Violin plots show the estimates under SW, IR, and PR; the dashed line is the simulated GTE. }
\label{fig:controlled_s1}
\end{figure}

\paragraph{Scenario 1: reducing downward mean bias.}
In Scenario~1, the control and treatment forecasts are
\[
\hat\mu_{n,t}(w)=\mu_n+\alpha_n\Delta(w),\qquad \hat\alpha_{n,t}=\alpha_n,
\]
where $\Delta(0)=-0.50$ and $\Delta(1)=-0.05$. Thus both groups underforecast the demand mean, but the treatment is less biased. Figure~\ref{fig:controlled_s1} reports the results under tight, medium, and loose capacity.


The results align with the Scenario~1 theory. The simulated GTE  is positive in all capacity regimes, consistent with Proposition~\ref{thm:S1_gte}.  Because reducing downward bias increases expected leftover inventory, the SW estimator is negatively biased, consistent with Theorem~\ref{thm:S1_sw}. Under tight capacity, IR substantially overestimates the GTE, reflecting the capacity-crowding channel in Theorem~\ref{thm:S1_IR}. As capacity becomes loose, the IR bias attenuates toward zero, which is consistent with the mechanism: when the shared constraint rarely binds, cross-item competition is weak. PR remains below IR in all regimes, consistent with Theorem~\ref{thm:S1_PR}. The results also illustrate that PR need not lie between SW and IR: when carryover dominates cross-sectional crowding, PR can become negatively biased.


\begin{figure}[t]
\centering
\begin{subfigure}{0.32\textwidth}
\centering
\includegraphics[width=\linewidth]{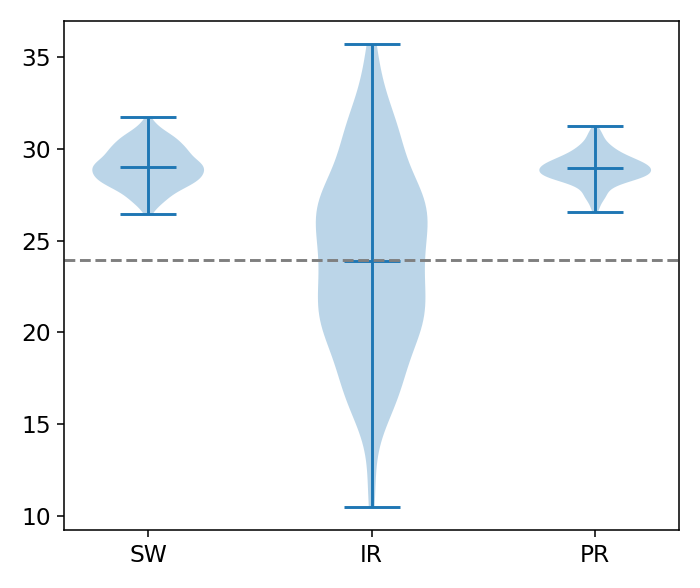}
\caption{Tight capacity}
\end{subfigure}\hfill
\begin{subfigure}{0.32\textwidth}
\centering
\includegraphics[width=\linewidth]{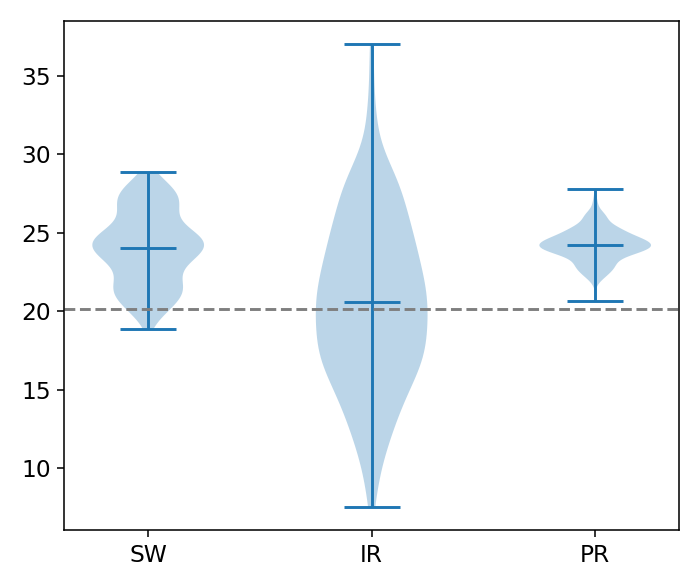}
\caption{Medium capacity}
\end{subfigure}\hfill
\begin{subfigure}{0.32\textwidth}
\centering
\includegraphics[width=\linewidth]{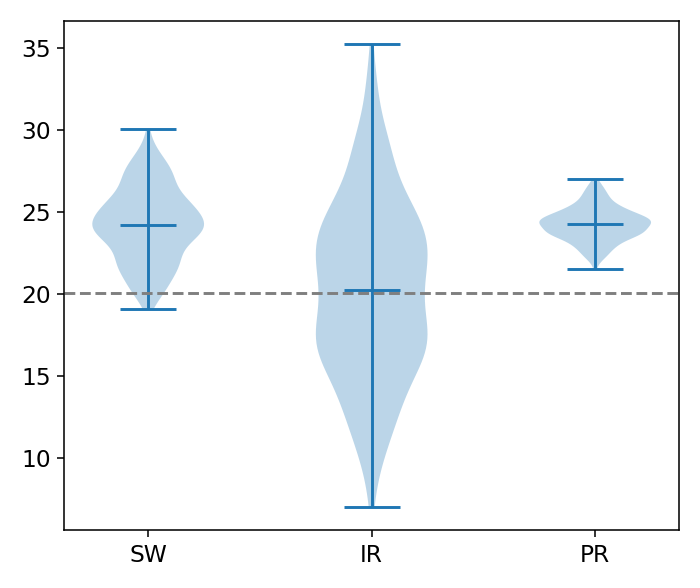}
\caption{Loose capacity}
\end{subfigure}
\caption{Simulation for Scenario~2. Treatment reduces forecast-error dispersion while preserving the mean. Violin plots show the estimates under SW, IR, and PR; the dashed line is the simulated GTE.}
\label{fig:controlled_s2}
\end{figure}

\paragraph{Scenario 2: reducing forecast-error dispersion.}
In Scenario~2, treatment and control forecasts have the same mean but different dispersion:
\[
\hat\mu_{n,t}(w)=\mu_n+\epsilon_{n,t}(w),\qquad \hat\alpha_{n,t}=\alpha_n,
\]
where $\epsilon_{n,t}(0)\sim \mathrm{Unif}[-30,30]$ and $\epsilon_{n,t}(1)\sim \mathrm{Unif}[-0.2,0.2]$. Hence $\mathbb E[\epsilon_{n,t}(w)]=0$ and $\epsilon_{n,t}(1)\le_{cx}\epsilon_{n,t}(0)$. We use a large system with $N=3000$, matching the mean-field nature of Section~4.2.  Figure~\ref{fig:controlled_s2} reports the results.



The Scenario~2 results match the large-$N$ theory. The GTE is positive, supporting Proposition~\ref{thm:S2_gte}: reducing forecast-error dispersion improves expected profit even when the forecast remains centered at the true mean. SW is upward biased, consistent with Theorem~\ref{thm:S2_sw}: the lower-dispersion treatment forecast produces less leftover inventory than control, so switchback carryover inflates the treatment-control contrast. IR is approximately unbiased in all three regimes, consistent with Theorem~\ref{thm:S2_ir}. PR is positive and very close to SW, consistent with Theorem~\ref{thm:S2_pr}. The effect of capacity is weaker than in Scenario~1 because treatment and control have the same mean; capacity changes the common multiplier but does not materially alter the relative dispersion advantage of treatment. 

\subsection{Trace-Driven Experiments on Fresh-Retail Data}\label{subsec:trace_driven_sim}

We next examine whether the same mechanisms appear in a more realistic trace-driven environment. We use FreshRetailNet-50K~\citep{wang2025freshretailnet}, which contains 50{,}000 store--product time series of hourly sales from 898 stores in 18 major cities, together with verified stockout annotations. Because sales are censored during stockouts, we use the latent-demand recovery model of \citet{wang2025freshretailnet} to construct daily recovered demand traces. Each forecasting model outputs a point forecast, and the simulator evaluates the resulting inventory policies on the recovered demand path.

The trace-driven experiment differs from the simulations above in two important ways. First, the demand path is fixed and recovered from real data rather than drawn from a uniform distribution. Second, we activate stockout substitution: when a product stocks out, a portion of unmet demand is redistributed to similar products in the same store. This module captures a real-world feature that is absent from the stylized theory but important in fresh retail. This is intended as mechanism validation; implementation details and the substitution construction are in Appendix~\ref{app:implementation_details}.

We use WPE to measure systematic forecast bias and WAPE to measure error magnitude:
\[
\mathrm{WAPE}:=\frac{\sum_{n,t}|\hat d_{n,t}-d_{n,t}|}{\sum_{n,t}d_{n,t}},\qquad
\mathrm{WPE}:=\frac{\sum_{n,t}(\hat d_{n,t}-d_{n,t})}{\sum_{n,t}d_{n,t}}.
\]
Because the trace-driven forecasts are point forecasts, WAPE is used as an
empirical proxy for forecast-error dispersion. Scenario~1 uses a control forecast with negative WPE and a treatment forecast with much smaller bias; Scenario~2 uses treatment and control forecasts with similar WPE but lower WAPE under treatment. The forecast pairs are summarized in Table~\ref{tab:fresh_forecast_pairs}. We compare forecasts generated by three forecasting methods: Naive, a simple
weekday-lag benchmark; DLinear~\citep{zeng2023transformers}; and
TFT~\citep{lim2021temporal}. Details of these forecasting methods are provided
in Appendix~\ref{appendix:forecasting-models}.

\begin{table}[t]
\centering
\caption{Forecast pairs used in the FreshRetailNet trace-driven experiments.}
\label{tab:fresh_forecast_pairs}
\begin{tabular}{llccclcc}
\toprule
Scenario & Treatment & WAPE & WPE & & Control & WAPE & WPE \\
\midrule
Scenario~1 & less-biased DLinear & 0.28 &  0.02 & & biased DLinear & 0.32 & -0.16 \\
Scenario~2 & TFT                 & 0.26 & -0.01 & & Naive          & 0.44 &  0.00 \\
\bottomrule
\end{tabular}
\end{table}

\begin{figure}[t] 
\centering \begin{subfigure}{0.32\linewidth} \centering \includegraphics[width=\linewidth]{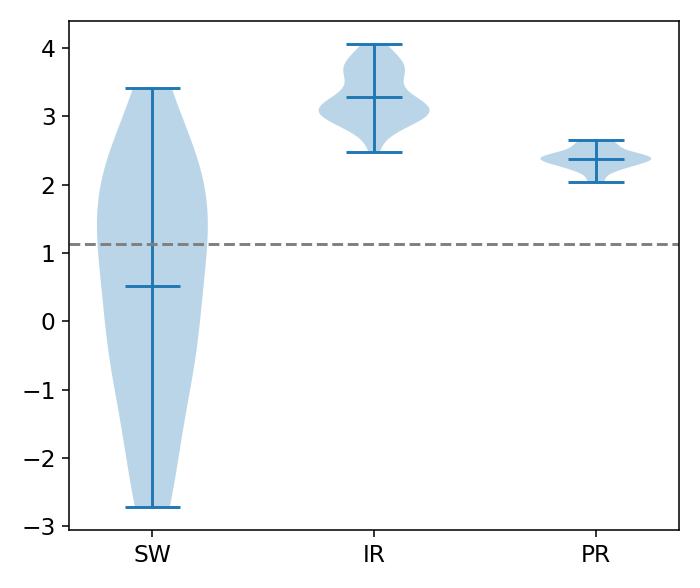} \caption{Tight capacity} \end{subfigure}\hfill \begin{subfigure}{0.32\linewidth} \centering \includegraphics[width=\linewidth]{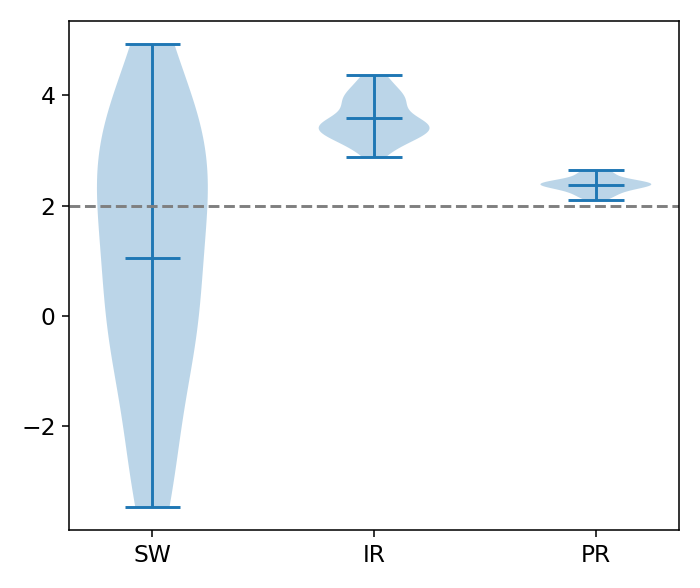} \caption{Medium capacity} \end{subfigure}\hfill \begin{subfigure}{0.32\linewidth} \centering \includegraphics[width=\linewidth]{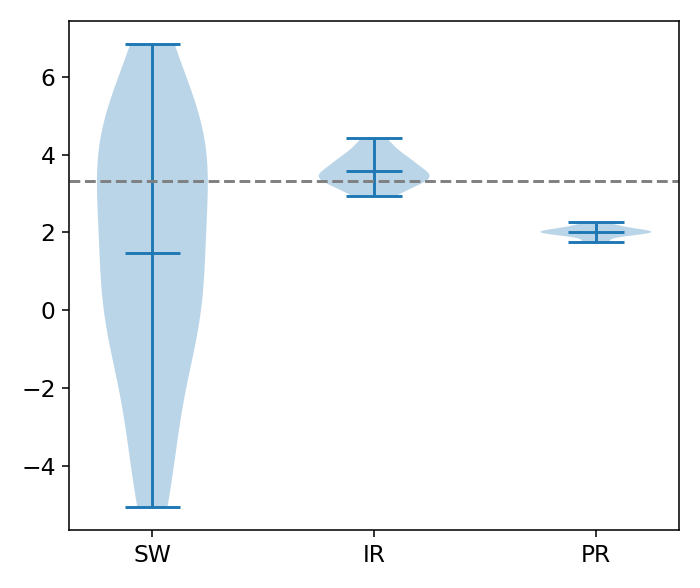} \caption{Loose capacity} \end{subfigure} \caption{Trace-driven FreshRetailNet experiments for Scenario~1 with stockout substitution.} \label{fig:fresh_s1_sub} \end{figure}

\paragraph{Scenario 1: biased demand mean.}
For Scenario~1, the control forecast is trained on censored sales and systematically underestimates recovered demand, while the treatment forecast is trained on recovered demand and is much less biased. Figure~\ref{fig:fresh_s1_sub} shows the results under different capacity levels.

The trace-driven results reproduce the Scenario~1 mechanisms in a noisier and more realistic environment. SW tends to underestimate the GTE and has much higher dispersion than IR and PR, mainly because the 7-day horizon provides limited temporal replication: SW randomizes by time blocks, whereas IR and PR average over many store--product series within each day. IR overestimates the GTE under tight and moderate capacity, reflecting the crowding channel in which treated items with higher targets compete for shared capacity and reduce inventory available to control items. As capacity becomes loose, this channel weakens and IR moves closer to the GTE. PR combines the two forces: cross-sectional crowding pushes the estimate upward, while inventory carryover pushes it downward. Thus, PR often reduces IR's positive bias, but can fall below the GTE when carryover is strong. We further note that, in this numerical study, Assumptions~\ref{asp:no_overshoot} and ~\ref{asp:margin_bounded_multiplier} may not always hold, which further underscores the robustness of our insights.

\begin{figure}[!t] 
\centering 
\begin{subfigure}{0.32\linewidth} 
\centering \includegraphics[width=\linewidth]{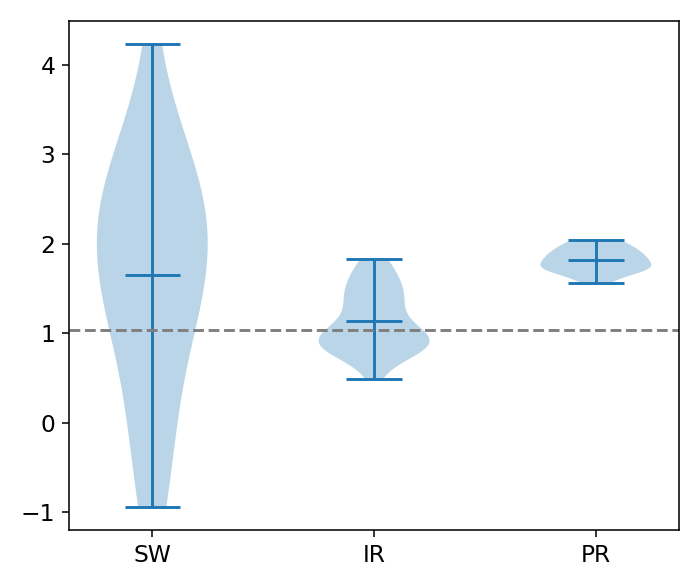} 
\caption{Tight capacity} 
\end{subfigure}
\hfill 
\begin{subfigure}{0.32\linewidth} 
\centering \includegraphics[width=\linewidth]{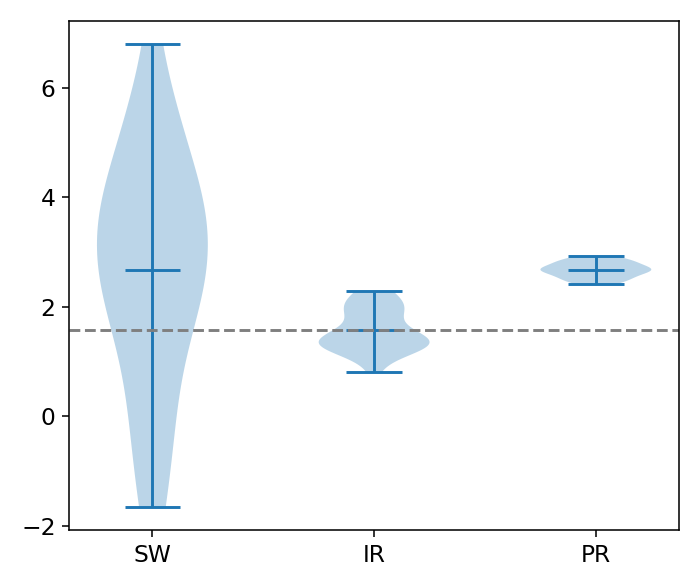} \caption{Medium capacity} 
\end{subfigure}
\hfill 
\begin{subfigure}{0.32\linewidth} 
\centering \includegraphics[width=\linewidth]{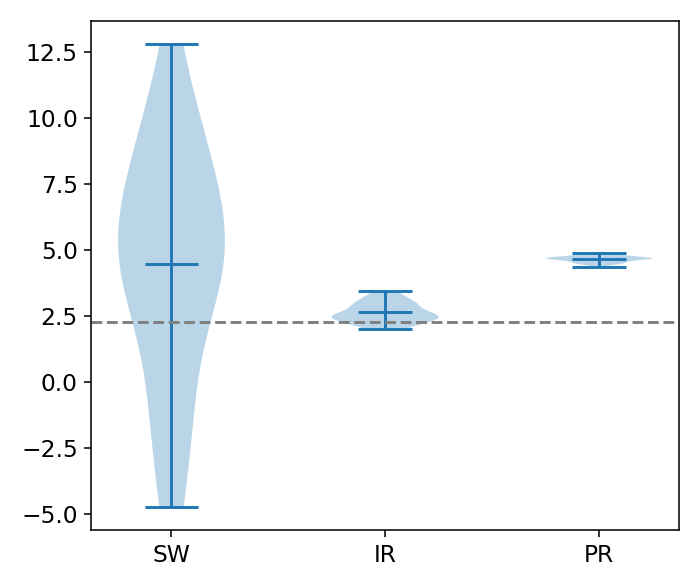} \caption{Loose capacity} 
\end{subfigure} \caption{Trace-driven FreshRetailNet experiments for Scenario~2 with stockout substitution.} \label{fig:fresh_s2_sub} \end{figure}

\paragraph{Scenario 2: lower forecast-error dispersion.}
For Scenario~2, treatment and control have similar WPE, while the treatment forecast has substantially lower WAPE. Figure~\ref{fig:fresh_s2_sub} reports the substitution-enabled A/B tests.

The treatment effect remains positive after stockout substitution is introduced. IR remains close to the GTE, while SW and PR are upward biased, matching the carryover mechanism in the controlled stochastic simulations and in Theorems~\ref{thm:S2_sw} and~\ref{thm:S2_pr}. Capacity has a weaker visible effect than in Scenario~1, which is consistent with the mean-field intuition that, when treatment and control have the same mean, capacity primarily shifts a common multiplier rather than changing the relative dispersion advantage of the treatment forecast.


\section{Conclusion}\label{section:conclusion}

In this paper, we study A/B testing strategies in multi-item, multi-period inventory systems with lost sales and capacity constraints. When the demand estimators used by the treatment and control primarily differ in their means, we show that switchback designs systematically underestimate, whereas item-level randomization systematically overestimates, the global treatment effect. When the demand estimators primarily differ in their forecast-error dispersion, switchback designs typically overestimate, while item-level randomization is asymptotically unbiased. Finally, we show that pairwise randomization can serve as a superior alternative in certain regimes. Based on our theoretical and numerical findings, we offer the following practical recommendations:

\begin{itemize}
    \item \textbf{Item-level randomization (IR).} Use IR when the capacity constraint is loose or when the treatment and control have similar estimation bias but differ in forecast-error dispersion. In these regimes, IR exhibits relatively low bias and lower dispersion.
    \item \textbf{Pairwise randomization (PR).} Use PR when the treatment and control differ substantially in estimation bias and the capacity constraint is relatively tight. In these scenarios, PR achieves lower bias and sampling variability than both IR and switchback designs.
     \item \textbf{Be cautious with switchback designs (SW).} In multi-period inventory systems, switchback experiments can suffer from systematic bias due to inventory carryover across periods and typically exhibit relatively high sampling variability. We therefore caution against their use in such settings.
\end{itemize}

\section*{Acknowledgement}

N. Si's research is partially funded by the Young Scientists Fund - Category C
(Grant No. 72501243) from the Natural Science Foundation of China
(NSFC), HKUST Li \& Fung Supply Chain Institute Research Grant 2025, and  the Hong Kong Research Grants Council [Theme-based Research Scheme T32-615/24-R].

\bibliographystyle{ACM-Reference-Format}
\bibliography{reference}

\appendix

\section{Proofs for Section~\ref{section:model} and Section~\ref{section:scenario1}}
\label{appendix:scenario1-proofs}

This appendix first derives two algebraic ingredients used throughout Scenario~1: the uniform-demand expression for expected leftover inventory and the affine response of base-stock levels under the margin assumptions. We then use these ingredients to prove the sign of GTE and the biases of SW, IR, and PR.

Here we record the form of the expected leftover under uniform demand.
For $D_{n,t}\sim{\rm Unif}[\mu_{n,t}-\alpha_{n,t},\mu_{n,t}+\alpha_{n,t}]$, write
$S=\mu_{n,t}+\alpha_{n,t}z$. A direct calculation gives
\[
\EE(S-D_{n,t})^+
=
\begin{cases}
0, & z\le -1,\\[0.3em]
\dfrac{\alpha_{n,t}}{4}(z+1)^2, & -1<z<1,\\[0.6em]
\alpha_{n,t}z, & z\ge 1.
\end{cases}
\]
We define
\[
z_{n,t}^{(\Delta)}
:=\frac{\hat S_{n,t}^{(\Delta)}-\mu_{n,t}}{\alpha_{n,t}}
= \Delta + 2\phi_{n,t}(\Delta)-1,\qquad
\phi_{n,t}(\Delta)
:=\bigg(\frac{m_n-\hat\lambda_t^{(\Delta)}}{M_n}\bigg)\in[0,1].
\]
Hence, for any $\Delta\le 0$,
\[
z_{n,t}^{(\Delta)}\le \Delta+1\le 1,
\]
so we never have $z_{n,t}^{(\Delta)}\ge 1$ in the range of interest $\Delta\le0$.
Thus only the cases $z_{n,t}^{(\Delta)}\le -1$ and $-1<z_{n,t}^{(\Delta)}<1$ are
relevant for our analysis. Lemma~\ref{lem:Delta0} precisely rules out, on a
common interval $(\underline{\Delta},0]$, the partially binding situation where
$\hat\lambda_t^{(\Delta)}>0$ and $z_{n,t}^{(\Delta)}\le -1$ for some~$n$.

\subsection{Proof of Proposition~\ref{prop:policy}}
Fix a period $t$. Recall that
\[
R_{n,t}^+(s,D_{n,t})=m_n s-M_n(s-D_{n,t})^+.
\]
The myopic problem is
\[
\max_{S_{1,t},\ldots,S_{N,t}}
\sum_{n=1}^N \EE R_{n,t}^+(S_{n,t},D_{n,t})
\quad
\text{s.t.}\quad
S_{n,t}\ge I_{n,t},\ \forall n,\qquad
\sum_{n=1}^N S_{n,t}\le B.
\]
For a Lagrange multiplier $\lambda\ge0$ associated with the capacity constraint,
the Lagrangian can be written as
\[
\mathcal L(S,\lambda)
=
\sum_{n=1}^N
\left\{
\EE R_{n,t}^+(S_{n,t},D_{n,t})-\lambda S_{n,t}
\right\}
+\lambda B.
\]
Thus, for a fixed $\lambda$, the item-level problem is
\[
\max_{s\ge I_{n,t}}\ \ell_{n,t}(s;\lambda),
\qquad
\ell_{n,t}(s;\lambda)
:=
(m_n-\lambda)s-M_n\EE(s-D_{n,t})^+.
\]
Since $s\mapsto \EE(s-D_{n,t})^+$ is convex, $\ell_{n,t}(s;\lambda)$ is concave in $s$.
Therefore, the following first-order/subgradient characterization is sufficient.

First consider the case $0\le\lambda<m_n$. In this case, the unconstrained
maximizer solves
\[
0=(m_n-\lambda)-M_nF_{n,t}(s),
\]
or equivalently
\[
F_{n,t}(s)=\frac{m_n-\lambda}{M_n}.
\]
Hence the unconstrained newsvendor target is
\[
F_{n,t}^{-1}\left(\frac{m_n-\lambda}{M_n}\right).
\]
After imposing the lower bound $s\ge I_{n,t}$, the item-level maximizer is
\[
\tilde S_{n,t}(\lambda)
=
\max\left\{
I_{n,t},\,
F_{n,t}^{-1}\left(\frac{m_n-\lambda}{M_n}\right)
\right\}.
\]

Next consider the case $\lambda>m_n$. For any $s\ge I_{n,t}$,
\begin{align*}
\ell_{n,t}(s;\lambda)-\ell_{n,t}(I_{n,t};\lambda)
&=(m_n-\lambda)(s-I_{n,t})
-M_n\EE\left[(s-D_{n,t})^+-(I_{n,t}-D_{n,t})^+\right].
\end{align*}
Since $s\ge I_{n,t}$, the map $x\mapsto (x-D_{n,t})^+$ is nondecreasing, so
\[
\EE\left[(s-D_{n,t})^+-(I_{n,t}-D_{n,t})^+\right]\ge0.
\]
Moreover, $m_n-\lambda<0$. Therefore, for every $s>I_{n,t}$,
\[
\ell_{n,t}(s;\lambda)-\ell_{n,t}(I_{n,t};\lambda)<0.
\]
Thus the unique item-level maximizer is
\[
\tilde S_{n,t}(\lambda)=I_{n,t}.
\]

Finally, consider the case $\lambda=m_n$. Then, for any $s\ge I_{n,t}$,
\[
\ell_{n,t}(s;m_n)-\ell_{n,t}(I_{n,t};m_n)
=
-M_n\EE\left[(s-D_{n,t})^+-(I_{n,t}-D_{n,t})^+\right]\le0.
\]
Let $L_{n,t}:=F_{n,t}^{-1}(0)$ denote the lower support of $D_{n,t}$. If $I_{n,t}\le L_{n,t}$,
then for every $s\in[I_{n,t},L_{n,t}]$, we have $(s-D_{n,t})^+=(I_{n,t}-D_{n,t})^+=0$ almost surely,
and hence all such $s$ are maximizers. If $I_{n,t}>L_{n,t}$, then the lower-bound
constraint already places the item inside the support, and the unique maximizer
is $s=I_{n,t}$. Therefore the maximizer set at $\lambda=m_n$ is
\[
\left[I_{n,t},\max\{I_{n,t},F_{n,t}^{-1}(0)\}\right].
\]

It remains to characterize the multiplier. Since the objective is concave in
$S$ and the feasible region is convex, the KKT conditions are sufficient for
optimality. Therefore an optimal multiplier $\lambda^*\ge0$ and item-level
maximizers $S_{n,t}^*(\lambda^*)$ must satisfy primal feasibility and complementary
slackness:
\[
\sum_{n=1}^N S_{n,t}^*(\lambda^*)\le B,
\qquad
\lambda^*\left(B-\sum_{n=1}^N S_{n,t}^*(\lambda^*)\right)=0.
\]
Equivalently, if the capacity constraint is slack under the unconstrained
item-level maximizers, then $\lambda^*=0$; otherwise $\lambda^*>0$ is chosen
so that the capacity constraint binds, with any necessary selection from the
maximizer set at $\lambda^*=m_n$. 
\qed

\subsection{Proof of Lemma~\ref{lem:Delta0}}
\label{app:Delta0-construction}

We first prove the following lemma regarding the base-stock level.

\begin{lemma}[Affine response under no overshoot]
\label{lem:s1_affine_response}
Consider Scenario~1 under Assumptions~\ref{asp:no_overshoot} and ~\ref{asp:margin_bounded_multiplier} . Since the Scenario~1 forecasts are weakly below the true demand mean, the induced estimated multipliers are also margin-bounded. For every assignment realization considered in the analysis, the inventory lower bound in~\eqref{eq:estimated} is inactive in every period. In particular,
\begin{align}
    \hat S_{n,t}=\bar S_{n,t}
    =\hat\mu_{n,t}(W_{n,t})+
\alpha_{n,t}\left(2\frac{m_n-\hat\lambda_t(\W_t)}{M_n}-1\right),
    \qquad \forall n\in[N],\ t\in[T].
    \label{eq:Slevel_under_asp1}
\end{align}
\end{lemma}

\emph{Proof.}
Fix an arbitrary assignment realization and forecast realization satisfying Assumption~\ref{asp:no_overshoot}. Under the inherited margin bound implied by Assumption~\ref{asp:margin_bounded_multiplier} in Scenario~1, the exact estimated response in~\eqref{eq:estimated} is
\[
\hat S_{n,t}=\max\{I_{n,t},\bar S_{n,t}\}.
\]
Moreover, because $\hat\lambda_t(\W_t)<m_n$ and $\hat\alpha_{n,t}=\alpha_{n,t}$, we have
\[
\bar S_{n,t}
=
\hat\mu_{n,t}(W_{n,t})+
\alpha_{n,t}\left(2\frac{m_n-\hat\lambda_t(\W_t)}{M_n}-1\right)
>
\hat\mu_{n,t}(W_{n,t})-\alpha_{n,t}\ge 0,
\]
where the last inequality follows from Assumption~\ref{asp:no_overshoot}(i). Hence $I_{n,1}=0\le \bar S_{n,1}$ and therefore $\hat S_{n,1}=\bar S_{n,1}$.

Suppose inductively that $I_{n,t}\le \bar S_{n,t}$, so that $\hat S_{n,t}=\bar S_{n,t}$. Since $D_{n,t}\ge \underline D_{n,t}$ and Assumption~\ref{asp:no_overshoot}(ii) gives
\[
\bar S_{n,t}-\bar S_{n,t+1}\le \underline D_{n,t},
\]
we obtain
\[
\bar S_{n,t}-D_{n,t}
\le
\bar S_{n,t}-\underline D_{n,t}
\le
\bar S_{n,t+1}.
\]
Because $\bar S_{n,t+1}\ge0$, it follows that
\begin{align}
I_{n,t+1}
=(\hat S_{n,t}-D_{n,t})^+
=(\bar S_{n,t}-D_{n,t})^+
\le \bar S_{n,t+1}.
\label{eq:s1:no-overshoot-sw}
\end{align}
Thus the inventory lower bound is inactive in period $t+1$. The claim follows by induction over $t$.
\qed

By Lemma~\ref{lem:s1_affine_response}, under Assumptions~\ref{asp:no_overshoot} and ~\ref{asp:margin_bounded_multiplier}  the actual base-stock level coincides with the affine target in every period. 

Recall the shorthand quantities $u_n$ and $v_n$ defined in Section~\ref{sec:dynamics}, and define, for each period $t$,
\[
A_{1,t}:=\sum_n 2\alpha_{n,t}v_n,\qquad
A_{2,t}:=\sum_n \alpha_{n,t},\qquad
C_t:=\sum_n\bigl[\mu_{n,t}+\alpha_{n,t}(2u_n-1)\bigr]-B.
\]
As we have discussed, 
under Assumption~\ref{asp:margin_bounded_multiplier}, we consider the following two cases:

\paragraph*{Case 1: $\lambda_t^*=0$ (weakly slack at $\Delta=0$).}
When the
capacity constraint is weakly slack, the targeted base-stock level becomes
\[
\hat S_{n,t}^{(\Delta)}
= \mu_{n,t}+\alpha_{n,t}\bigl(\Delta+2u_n-1\bigr),
\]
which is increasing in $\Delta$. Thus when $\Delta<0$, we have $\sum_n\hat S_{n,t}^{(\Delta)} < B$, and at $\Delta=0$ we have $\sum_n\hat S_{n,t}^{(0)}\le B$ by $\lambda_t^*=0$. Hence the system remains weakly slack in period $t$ for all $\Delta\le0$. To ensure a finite lower bound,  we set $\widetilde\Delta_t=-M$ for such $t$, where $M$ is a sufficiently large positive constant.

\paragraph*{Case 2: $0<\lambda_t^*<\min_{n\in[N]} m_n$ (strictly interior at $\Delta=0$).}
Because the solution satisfies
$0<\lambda_t^*<\min_{n\in[N]} m_n$,  for every $n\in[N]$ the fraction
\[
\frac{m_n-\lambda_t^*}{M_n}
\]
lies in $(0,1)$ and no truncation occurs. We would like to find a neighborhood of $\Delta=0$ such that
the solution remains interior, so the margin boundary is not reached and the lower-support boundary remains inactive.
In such an interior neighborhood we have
\[
\hat S_{n,t}^{(\Delta)}
= \mu_{n,t} + \alpha_{n,t}\bigl(\Delta + 2u_n -1 - 2v_n\hat\lambda_t^{(\Delta)}\bigr),
\]
and the capacity constraint binds:
\[
\sum_n \hat S_{n,t}^{(\Delta)} = B
\;\Longleftrightarrow\;
A_{2,t}\Delta + C_t - A_{1,t}\hat\lambda_t^{(\Delta)}=0.
\]
Hence, in the interior regime,
\begin{equation}
\label{eq:lambda-linear-appendix}
\hat\lambda_t^{(\Delta)} 
= \frac{A_{2,t}\Delta + C_t}{A_{1,t}},
\qquad
\hat\lambda_t^{(0)}=\lambda_t^*=\frac{C_t}{A_{1,t}}>0.
\end{equation}
The normalized order position is
\begin{align}
    \label{eq:z-linear-appendix}
    z_{n,t}^{(\Delta)} 
= \frac{\hat S_{n,t}^{(\Delta)}-\mu_{n,t}}{\alpha_{n,t}}
= \Delta + 2u_n -1 - 2v_n\hat\lambda_t^{(\Delta)}
= \beta_{n,t}\Delta + \gamma_{n,t},
\end{align}
where
\begin{equation}
\label{eq:beta-gamma-appendix}
\beta_{n,t} := 1 - \frac{2v_nA_{2,t}}{A_{1,t}},
\qquad
\gamma_{n,t} := 2u_n-1 - \frac{2v_nC_t}{A_{1,t}}.
\end{equation}
Thus, in the interior regime, both $\hat\lambda_t^{(\Delta)}$ and
$z_{n,t}^{(\Delta)}$ are affine and hence continuous in~$\Delta$.

Potentially, as we decrease $\Delta$ from $0$ to negative values, $\hat\lambda_t^{(\Delta)} $ decreases as well, and the solution
may leave the interior regime in two ways:
\begin{enumerate}
\item[(a)] the capacity becomes slack, i.e., $\hat\lambda_t^{(\Delta)}$ hits~$0$;
\item[(b)] for some $n$, the normalized order position hits the lower support, i.e.,
$z_{n,t}^{(\Delta)}=-1$.
\end{enumerate}
Using \eqref{eq:lambda-linear-appendix}, (a) occurs when
\[
\hat\lambda_t^{(\Delta)}=0
\iff A_{2,t}\Delta + C_t=0
\iff \Delta = \Delta_t^\lambda := -\frac{C_t}{A_{2,t}}<0,
\]
since $C_t=A_{1,t}\lambda_t^*>0$ and $A_{2,t}>0$. Using
$z_{n,t}^{(\Delta)}=\beta_{n,t}\Delta+\gamma_{n,t}$, (b) occurs for item $n$
when
\[
\beta_{n,t}\Delta + \gamma_{n,t}=-1
\iff \Delta = \Delta_{n,t}^z := \frac{-1-\gamma_{n,t}}{\beta_{n,t}},
\]
provided $\beta_{n,t}>0$. If $\beta_{n,t}\le0$, then $z_{n,t}^{(\Delta)}$
achieves its minimum over $[\Delta,0]$ at $\Delta=0$; under Assumption~\ref{asp:margin_bounded_multiplier}, $z_{n,t}^{(0)}>-1$, so no additional restriction
is required for this $n$, and we may set $\Delta_{n,t}^z:=-\infty$.

Define, for this strictly interior period~$t$,
\[
\widetilde\Delta_t
:= \max\Big\{\Delta_t^\lambda,\ \max_n \Delta_{n,t}^z\Big\}<0.
\]
By construction, for all $\Delta\in (\widetilde\Delta_t,0]$ either
\begin{itemize}
\item $\hat\lambda_t^{(\Delta)}=0$ (weakly slack capacity), or
\item $\hat\lambda_t^{(\Delta)}>0$ and $z_{n,t}^{(\Delta)}>-1$ for all $n$
(strictly interior).
\end{itemize}
In particular, the partially binding regime with $\hat\lambda_t^{(\Delta)}>0$
and $z_{n,t}^{(\Delta)}\le -1$ for some $n$ does not occur on
$(\widetilde\Delta_t,0]$.

\medskip

Finally, combining Case~1 and Case~2, for each period $t$ we have constructed
some $\widetilde\Delta_t<0$ such that on $(\widetilde\Delta_t,0]$ the
solution is either weakly slack or strictly interior in the sense of
Lemma~\ref{lem:Delta0}. Let
\[
\underline{\Delta}:=\max_t \widetilde\Delta_t<0.
\]
Then for all $t$ and all $\Delta\in(\underline{\Delta},0]$, the optimal
solution is either weakly slack or strictly interior, as claimed.
This completes the proof of Lemma~\ref{lem:Delta0}.

\subsection{Proofs of Proposition~\ref{thm:S1_gte} and Theorem~\ref{thm:S1_sw}}  \label{appendix:proof_s1_gte_sw}
\subsubsection{Derivation of the GTE and Bias Expression}
\label{app:bias-expression}

By the definition of $R_{n,t}^+$, we have
\begin{align}
     & \qquad \sum_{t=1}^T R_{n,t} \nonumber \\
     & = \sum_{t=1}^T \Big(R_{n,t}^+(I_{n,t} + O_{n,t},D_{n,t}) + c_nI_{n,t} - c_n(I_{n,t} + O_{n,t} - D_{n,t})^+  \Big) +c_n   (I_{n,T} + O_{n,T}-D_{n,T})^+ \nonumber \\
    & = \sum_{t=1}^T  R_{n,t}^+(\hat S_{n,t},D_{n,t}) + c_n  I_{n,1} -  c_n(\hat{S}_{n,1} - D_{n,1})^+  \nonumber\\
    &\quad +  \sum_{t=2}^T  c_n \Big[   (\hat{S}_{n,t-1} - D_{n,t-1})^+ - (\hat S_{n,t} - D_{n,t})^+ \Big]
    + c_n   (\hat{S}_{n,T}-D_{n,T})^+ \nonumber\\
    & = \sum_{t=1}^T  R_{n,t}^+(\hat S_{n,t},D_{n,t}) + c_n  I_{n,1} -  c_n(\hat{S}_{n,1} - D_{n,1})^+   + \left(  c_n     (\hat{S}_{n,1} - D_{n,1})^+ - c_n(\hat S_{n,T} - D_{n,T})^+ 
    \right)+ c_n   (\hat{S}_{n,T}-D_{n,T})^+ \nonumber\\
    & = \sum_{t=1}^T R_{n,t}^+(\hat S_{n,t} ,D_{n,t}) + c_nI_{n,1}. 
    \label{eq: sum_r_+}
\end{align}
The equality above is a telescoping identity: the intermediate leftover terms $-c_n(\hat S_{n,t}-D_{n,t})^+$ and $+c_n(\hat S_{n,t}-D_{n,t})^+$ from adjacent periods cancel, and the terminal leftover is exactly offset by the salvage term.

By Lemma~\ref{lem:s1_affine_response}, under Assumptions~\ref{asp:no_overshoot} and ~\ref{asp:margin_bounded_multiplier} , the base-stock level satisfies~\eqref{eq:Slevel_under_asp1} for every assignment realization. We denote this target by $\hat S_{n,t}(\W_t)$.

Combining ~\eqref{eq: sum_r_+} with  the definition of $GTE$, we obtain
\begin{align}
    GTE 
    &=  \frac{1}{NT}  \sum_{n=1}^N   \sum_{t=1}^{T}   \EE \Big[R_{n,t} \mid \W =  \boldsymbol{1}_{N\times T}  \Big] 
       - \frac{1}{NT}  \sum_{n=1}^N   \sum_{t=1}^{T}   \EE \Big[R_{n,t} \mid \W =  \boldsymbol{0}_{N\times T} \Big] \nonumber\\
    & = \frac{1}{NT}  \sum_{n=1}^N   \sum_{t=1}^{T} \EE R^+_{n,t}(\hat S_{n,t}(\boldsymbol{1}), D_{n,t}) 
      - \frac{1}{NT}  \sum_{n=1}^N   \sum_{t=1}^{T} \EE R^+_{n,t}(\hat S_{n,t}(\boldsymbol{0}), D_{n,t}) \label{eq:GTE}.
\end{align}

We denote 
\begin{align}
        GT :=  \frac{1}{NT}  \sum_{n=1}^N   \sum_{t=1}^{T} \EE R^+_{n,t}(\hat S_{n,t}(\boldsymbol{1}), D_{n,t}), 
    \qquad
    GC :=  \frac{1}{NT}  \sum_{n=1}^N   \sum_{t=1}^{T} \EE R^+_{n,t}(\hat S_{n,t}(\boldsymbol{0}), D_{n,t}),\label{eq: GT}
\end{align}
and the corresponding IPW estimators
\[
    \widehat{GT} = \frac{1}{NT} \sum_{n=1}^N \sum_{t=1}^T \frac{W_{n,t} R_{n,t}}{p},
    \qquad
    \widehat{GC} = \frac{1}{NT} \sum_{n=1}^N \sum_{t=1}^T \frac{(1-W_{n,t}) R_{n,t}}{1-p}.
\]

Since
\[
 \EE \widehat{GT} 
 = \EE\left[ \frac{W_{n,t}R_{n,t}}{p} \right] 
 = \frac{\PP(W_{n,t} = 1)}{p} \EE \Big[ R_{n,t} \mid W_{n,t} = 1\Big] 
 =  \EE \Big[ R_{n,t} \mid W_{n,t} = 1\Big],
\]
we have
\begin{align*}
    \EE\widehat{GTE} 
    & =  \frac{1}{NT}\sum_{n=1}^N \sum_{t=1}^T \EE\left[ \frac{W_{n,t} R_{n,t}}{p} - \frac{(1-W_{n,t}) R_{n,t}}{1-p} \right] \\
    & =  \frac{1}{NT}\sum_{n=1}^N \sum_{t=1}^T \Big[ \EE \big[ R_{n,t} \mid W_{n,t} = 1\big]
                                                - \EE \big[ R_{n,t} \mid W_{n,t} = 0\big]\Big].
\end{align*}

In the appendix, we use superscripts to distinguish treatment assignments in different experiments. In the switchback experiment, $W_{n,t}^{SW} = 1$ implies $\W_t^{SW} =  \boldsymbol{1}$. By the definition of $R_{n,t}$,
\begin{align}
     \sum_{t=1}^T   \EE \Big[ R_{n,t} \mid W_{n,t}^{SW} = 1\Big]
     &=  \sum_{t=1}^T\EE\Big[ R^+_{n,t}(\hat S_{n,t}(\W_t^{SW}),D_{n,t})  + c_n  I_{n,t} - c_n (\hat S_{n,t}(\W_t^{SW}) - D_{n,t})^+ \,\bigm|\, W_{n,t}^{SW} = 1 \Big] \notag\\
     &\quad + c_n \EE\Big[(\hat S_{n,T}(\W_T^{SW}) - D_{n,T})^+ \,\bigm|\, W_{n,T}^{SW} = 1\Big] \notag\\
     &= \sum_{t=1}^T\EE R^+_{n,t}(\hat S_{n,t}(\boldsymbol{1}),D_{n,t}) + \sum_{t=1}^T c_n\EE \Big[I_{n,t} \mid W_{n,t}^{SW} = 1\Big]   - \sum_{t=1}^{T-1} c_n \EE(\hat S_{n,t}(\boldsymbol{1}) - D_{n,t})^+.
     \label{eq: pfthm1.1}
\end{align}
For $t \geqslant 2$, we have
\begin{align}
    \EE \Big[I_{n,t} \mid W_{n,t}^{SW} = 1\Big] 
    &= \EE \Big[(\hat S_{n,t-1}(\W_{t-1}^{SW}) - D_{n,t-1})^+ \mid W_{n,t}^{SW} = 1\Big]  \notag \\
    & = \PP\big[\W_{t-1}^{SW} = \boldsymbol{1} \mid W_{n,t}^{SW} = 1 \big] \EE(\hat S_{n,t-1}(\boldsymbol{1}) - D_{n,t-1})^+  \notag\\
    &\quad +\PP\big[\W_{t-1}^{SW} = \boldsymbol{0}  \mid W_{n,t}^{SW} = 1 \big] \EE(\hat S_{n,t-1}(\boldsymbol{0}) - D_{n,t-1})^+   \notag\\
    &= p \EE(\hat S_{n,t-1}(\boldsymbol{1}) - D_{n,t-1})^+  + (1-p) \EE (\hat S_{n,t-1}(\boldsymbol{0}) - D_{n,t-1})^+ , 
    \label{eq: pfthm1.2}
\end{align}
where the last equality uses $\W_{t-1}^{SW} \perp W_{n,t}^{SW}$. Substituting
\eqref{eq: pfthm1.2} into \eqref{eq: pfthm1.1}, we obtain
\begin{align*}
\quad &  \sum_{t=1}^T   \EE \Big[ R_{n,t} \mid W_{n,t}^{SW} = 1\Big] \\
 &=  \sum_{t=1}^T\EE R^+_{n,t}(\hat S_{n,t} (\boldsymbol{1}),D_{n,t}) + c_n I_{n,1} - \sum_{t=1}^{T-1} c_n(1-p) \Big[ \EE (\hat S_{n,t}(\boldsymbol{1}) - D_{n,t})^+  -  \EE (\hat S_{n,t}(\boldsymbol{0}) - D_{n,t})^+ 
 \Big].
\end{align*}

Since $I_{n,1} = 0$ for all $n$, combining with the definition of $GT$ yields
\begin{align*}
    \EE \widehat{GT}^{SW} - GT 
    & = \frac{1}{NT} \sum_{n=1}^N \sum_{t=1}^T   \EE \Big[ R_{n,t} \mid W_{n,t}^{SW} = 1\Big] 
    - \frac{1}{NT} \sum_{n=1}^N \sum_{t=1}^T \EE R_{n,t}^+(\hat S_{n,t}(\boldsymbol{1}),D_{n,t}) \\
    & = - \frac{1}{NT} \sum_{n=1}^N \sum_{t=1}^{T-1}  c_n(1-p) \Big[ \EE (\hat S_{n,t}(\boldsymbol{1}) - D_{n,t})^+ - \EE (\hat S_{n,t}(\boldsymbol{0}) - D_{n,t})^+ \Big].
\end{align*}

Similarly, for the control group,
\begin{align*}
       \EE \widehat{GC}^{SW} - GC 
       & =   \frac{1}{NT} \sum_{n=1}^N \sum_{t=1}^{T-1} c_n p\Big[ \EE (\hat S_{n,t}(\boldsymbol{1}) - D_{n,t})^+ - \EE (\hat S_{n,t}(\boldsymbol{0}) - D_{n,t})^+ \Big]. 
\end{align*}

Therefore, under Assumption~\ref{asp:no_overshoot}, we obtain
\begin{align*}
       \EE  \widehat{GTE}^{SW}- GTE 
       & =\big(\EE \widehat{GT}^{SW} - GT \big) -  \big(\EE \widehat{GC}^{SW} - GC \big) \\
       & =  - \frac{1}{NT} \sum_{n=1}^N \sum_{t=1}^{T-1}  c_n  \Big[ \EE (\hat S_{n,t}(\boldsymbol{1}) - D_{n,t})^+ - \EE (\hat S_{n,t}(\boldsymbol{0}) - D_{n,t})^+ \Big].
\end{align*}

\vspace{10pt}

Recall that $D_{n,t}\sim\mathrm{Unif}[\mu_{n,t}-\alpha_{n,t},\mu_{n,t}+\alpha_{n,t}]$ and
\[
\hat\mu_{n,t}^{(\Delta)}=\mu_{n,t}+\Delta\alpha_{n,t},\qquad \Delta\le 0.
\]
To facilitate the analysis, when all items have the same forecast error parameter $\Delta$ at time $t$, we add the superscript notation $^{(\Delta)}$ to $\hat{\lambda}_{t}$ and $\hat{S}_{n,t}$: 
\begin{align*}
 & \hat \lambda_t^{(\Delta)} = \max \left(0,\frac{\sum_{n=1}^N[\hat \mu_{n,t}^{(\Delta)} +\alpha_{n,t}(2u_n - 1)]-B}{\sum_{n=1}^N2\alpha_{n,t} v_n} \right),\\
     & \hat{S}_{n,t}^{(\Delta)}  = \hat \mu_{n,t}^{(\Delta)} + \Big(2 \Big(\frac{m_n-\hat \lambda_t^{(\Delta)}}{M_n}\Big) - 1 \Big) \alpha_{n,t},
\end{align*}
where $u_n$ and $v_n$ are defined in Section~\ref{sec:dynamics}.

We define $f_1$ and $f_2$ as
\begin{align}
    & f_1(\Delta)=\sum_{t\in[T]}\sum_{n\in[N]}\EE R^+_{n,t}(\hat S_{n,t}^{(\Delta)},D_{n,t})=\sum_{t\in[T]}\sum_{n\in[N]}\Big[m_n\hat S_{n,t}^{(\Delta)}-M_n\,\mathbb{E}(\hat S_{n,t}^{(\Delta)}-D_{n,t})^+\Big], \label{eq:def_f1}\\
    & f_2(\Delta)= -\sum_{t\in[T-1]}\sum_{n\in[N]} c_n\,\mathbb{E}(\hat S_{n,t}^{(\Delta)}-D_{n,t})^+,\label{eq:def_f2}
\end{align}
{By the definition of $f_1$ and $f_2$, 
we have
\[
GTE = \frac{1}{NT}f_1(\Delta_1) - \frac{1}{NT} f_1(\Delta_0),
\qquad
\EE  \widehat{GTE}^{SW}- GTE =  \frac{1}{NT}f_2(\Delta_1) -  \frac{1}{NT}f_2(\Delta_0).
\]
To obtain the sign of these expressions, it remains to analyze the monotonicity of
$f_1(\Delta)$ and $f_2(\Delta)$.
}

\subsubsection{Monotonicity of $f_1(\Delta)$}

For $t \in [T]$, define
\[
f_{1,t}(\Delta)
= \sum_{n}\Big[m_n\hat S_{n,t}^{(\Delta)}-M_n\,\mathbb{E}(\hat S_{n,t}^{(\Delta)}-D_{n,t})^+\Big].
\]

On the interval $\Delta\in(\underline{\Delta},0]$, each period $t$ is either weakly slack or strictly interior by Lemma~\ref{lem:Delta0}. We consider these two cases.

\begin{itemize}
    \item \emph{Slack regime} ($\hat\lambda_t^{(\Delta)}\equiv0$):

    In the weakly slack regime, we have
    \[
    \hat S_{n,t}^{(\Delta)} 
    = \mu_{n,t}+\alpha_{n,t}\bigl(\Delta+2u_n-1\bigr),
    \qquad
    z_{n,t}^{(\Delta)}=\Delta+2u_n-1.
    \]
    The contribution of item $(n,t)$ to $f_1$ is
    \[
    g_{n,t}(\Delta)
    := m_n\hat S_{n,t}^{(\Delta)}
       -M_n \EE(\hat S_{n,t}^{(\Delta)}-D_{n,t})^+.
    \]

    If $z_{n,t}^{(\Delta)}\le -1$, then $\EE(\hat S_{n,t}^{(\Delta)}-D_{n,t})^+=0$ and hence
    \[
    \frac{dg_{n,t}(\Delta)}{d\Delta}
    = m_n\frac{d\hat S_{n,t}^{(\Delta)}}{d\Delta}
    = m_n\alpha_{n,t}>0.
    \]

    If $-1<z_{n,t}^{(\Delta)}<1$, then
    \[
    \EE(\hat S_{n,t}^{(\Delta)}-D_{n,t})^+
    =\frac{\alpha_{n,t}}{4}\bigl(z_{n,t}^{(\Delta)}+1\bigr)^2,
    \]
    so that
    \[
    \frac{dg_{n,t}(\Delta)}{d\Delta}
    = m_n\alpha_{n,t}
    - M_n\cdot \frac{\alpha_{n,t}}{2}\bigl(z_{n,t}^{(\Delta)}+1\bigr)
    = \alpha_{n,t}\Big[m_n - \frac{M_n}{2}\bigl(z_{n,t}^{(\Delta)}+1\bigr)\Big].
    \]
    Using $z_{n,t}^{(\Delta)}+1=\Delta+2u_n$ and the fact that $\Delta\le 0$,
    \[
    z_{n,t}^{(\Delta)}+1 = \Delta+2u_n \le 2u_n.
    \]
    Since the coefficient in front of $(z_{n,t}^{(\Delta)}+1)$ is negative,
    the worst (smallest) value of the bracket is attained at the largest possible value of $z_{n,t}^{(\Delta)}+1$, namely $2u_n$ (corresponding to $\Delta=0$). Thus
    \[
    m_n - \frac{M_n}{2}\bigl(z_{n,t}^{(\Delta)}+1\bigr)
    \ge m_n - M_n u_n.
    \]
    With the definition $u_n = \frac{m_n}{M_n}$, we have   $ m_n - M_n u_n = 0$.
    
    Hence in all cases
    \[
    \frac{dg_{n,t}(\Delta)}{d\Delta} \ge 0,
    \]
    and summing over $n$ shows that $f_{1,t}(\Delta)$ is nondecreasing in $\Delta$ in the weakly slack regime.

    \item \emph{Strictly interior regime}:

    In the interior regime, $z_{n,t}^{(\Delta)}\in(-1,1)$ for all $n$, and
    \[
    \mathbb{E}(\hat S_{n,t}^{(\Delta)}-D_{n,t})^+
    =\frac{\alpha_{n,t}}{4}\bigl(z_{n,t}^{(\Delta)}+1\bigr)^2.
    \]
    Using the linear representations \eqref{eq:lambda-linear-appendix} and \eqref{eq:z-linear-appendix}, one can show that
    \[
    f_{1,t}'(\Delta)
    = \frac{\Delta}{A_{1,t}}\Big(A_{2,t}^2 - \frac{A_{1,t}}{2}G_t\Big),
    \]
    where
    \[
    G_t:=\sum_n \frac{\alpha_{n,t}}{v_n}=\sum_n \alpha_{n,t}M_n.
    \]
    Applying the Cauchy--Schwarz inequality, we obtain
    \[
    A_{2,t}^2 = \Big(\sum_n \alpha_{n,t}\Big)^2
    \le \Big(\sum_n \alpha_{n,t}v_n\Big)\Big(\sum_n \frac{\alpha_{n,t}}{v_n}\Big)
    = \frac{A_{1,t}}{2}G_t.
    \]
    Thus the bracket in $f_{1,t}'(\Delta)$ is nonpositive, and since $\Delta\le0$ and $A_{1,t}>0$, we have
    \[
    f_{1,t}'(\Delta)\ge0
    \qquad\text{for all }\Delta\in(\widetilde\Delta_t,0]
    \]
    whenever period $t$ is strictly interior.
\end{itemize}

Combining the above two regimes, we conclude that $f_{1,t}'(\Delta)\ge0$ on $(\underline{\Delta},0]$, hence
\[
f_1'(\Delta)=\sum_t f_{1,t}'(\Delta)\ge0,
\quad
\forall\,\Delta\in(\underline{\Delta},0],
\]
which implies that $f_1(\Delta)$ is non-decreasing on $(\underline{\Delta},0]$. The non-negative GTE follows from
\[
GTE = \frac{1}{NT} f_1(\Delta_1) - \frac{1}{NT} f_1(\Delta_0).
\]

\subsubsection{Monotonicity of $f_2(\Delta)$}

For $t\in [T-1]$, define 
\[
f_{2,t}(\Delta)= - \sum_{n} c_n\,\mathbb{E}(\hat S_{n,t}^{(\Delta)}-D_{n,t})^+.
\]

\begin{itemize}
\item \emph{Slack regime} ($\hat\lambda_t^{(\Delta)}\equiv0$):
As above, in the weakly slack regime, we have
\[
\hat S_{n,t}^{(\Delta)} 
= \mu_{n,t}+\alpha_{n,t}\bigl(\Delta+2u_n-1\bigr),
\qquad
z_{n,t}^{(\Delta)}=\Delta+2u_n-1.
\]
The contribution of item $(n,t)$ to $f_2$ is
\[
h_{n,t}(\Delta)
:= -\,c_n\,\EE\bigl(\hat S_{n,t}^{(\Delta)}-D_{n,t}\bigr)^+.
\]
On $\Delta\le0$ we only encounter the two cases $z_{n,t}^{(\Delta)}\le-1$ and $-1<z_{n,t}^{(\Delta)}<1$.

If $z_{n,t}^{(\Delta)}\le -1$, then $\EE(\hat S_{n,t}^{(\Delta)}-D_{n,t})^+=0$
and hence
\[
\frac{dh_{n,t}(\Delta)}{d\Delta}=0.
\]

If $-1<z_{n,t}^{(\Delta)}<1$, then
\[
\EE(\hat S_{n,t}^{(\Delta)}-D_{n,t})^+
=\frac{\alpha_{n,t}}{4}\bigl(z_{n,t}^{(\Delta)}+1\bigr)^2,
\]
so, using $\frac{dz_{n,t}^{(\Delta)}}{d\Delta}=1$,
\[
\frac{d}{d\Delta}\EE(\hat S_{n,t}^{(\Delta)}-D_{n,t})^+
= \frac{\alpha_{n,t}}{2}\bigl(z_{n,t}^{(\Delta)}+1\bigr),
\]
and therefore
\[
\frac{dh_{n,t}(\Delta)}{d\Delta}
= -\,c_n\,\frac{\alpha_{n,t}}{2}\bigl(z_{n,t}^{(\Delta)}+1\bigr).
\]
The regime $-1<z_{n,t}^{(\Delta)}<1$ implies $z_{n,t}^{(\Delta)}+1>0$, so
\[
\frac{dh_{n,t}(\Delta)}{d\Delta}\le 0.
\]

Combining the two cases, we obtain $\frac{dh_{n,t}(\Delta)}{d\Delta}\le0$ for
all $\Delta\le0$ in the weakly slack regime. Summing over $n$ shows that
$f_{2,t}(\Delta)=\sum_n h_{n,t}(\Delta)$ is non-increasing in $\Delta$, and
hence $f_2(\Delta)$ is non-increasing in $\Delta$.

\item \emph{Strictly interior regime}: We again have
\[
\mathbb{E}(\hat S_{n,t}^{(\Delta)}-D_{n,t})^+
=\frac{\alpha_{n,t}}{4}\bigl(z_{n,t}^{(\Delta)}+1\bigr)^2,
\quad
z_{n,t}^{(\Delta)}=\beta_{n,t}\Delta+\gamma_{n,t},
\]
with $\beta_{n,t},\gamma_{n,t}$ given by \eqref{eq:beta-gamma-appendix}. Thus
\[
f_{2,t}(\Delta)
= -\sum_n c_n\,\frac{\alpha_{n,t}}{4}\bigl(z_{n,t}^{(\Delta)}+1\bigr)^2.
\]
Differentiating with respect to $\Delta$ and using $\frac{dz_{n,t}^{(\Delta)}}{d\Delta}=\beta_{n,t}$, we obtain
\[
f_{2,t}'(\Delta)
= -\frac12\sum_n c_n\alpha_{n,t}\bigl(z_{n,t}^{(\Delta)}+1\bigr)\beta_{n,t}
= -\frac12\sum_n c_n\alpha_{n,t}\bigl(\beta_{n,t}\Delta+\gamma_{n,t}+1\bigr)\beta_{n,t}.
\]
Therefore
\[
f_{2,t}'(\Delta) = \theta_t\Delta + \kappa_t,
\]
where
\begin{align}\label{eq:at-bt-appendix}
   \theta_t := -\frac12\sum_n c_n\alpha_{n,t}\beta_{n,t}^2\le 0,
\qquad
\kappa_t := -\frac12\sum_n c_n\alpha_{n,t}\beta_{n,t}(\gamma_{n,t}+1). 
\end{align}
Hence, in the interior regime, $f_{2,t}'(\Delta)$ is an affine function of~$\Delta$ with nonpositive slope $\theta_t$.

On the interval $(\widetilde\Delta_t,0]$, period $t$ remains either weakly slack or strictly interior by Lemma~\ref{lem:Delta0}. In the weakly slack regime we already have $f_{2,t}'(\Delta)\le0$; in the interior regime, since $\theta_t\le0$, the function $\Delta\mapsto f_{2,t}'(\Delta)$ is nonincreasing. Its supremum over the open-left interval $(\widetilde\Delta_t,0]$ is approached as $\Delta\downarrow\widetilde\Delta_t$ and is bounded by
\[
\sup_{\Delta\in(\widetilde\Delta_t,0]} f_{2,t}'(\Delta)
\le \theta_t\widetilde\Delta_t + \kappa_t.
\]
Therefore, the explicit condition
\[
\theta_t\widetilde\Delta_t + \kappa_t \le 0
\]
is sufficient to guarantee $f_{2,t}'(\Delta)\le0$ for all $\Delta\in(\widetilde\Delta_t,0]$ for that period $t$.
\end{itemize}

In conclusion,
under Assumption~\ref{asp:margin_bounded_multiplier}, if, in addition, $\theta_t\widetilde\Delta_t+\kappa_t\le0$ for every period $t$ that is strictly interior, then
\[
f_{2,t}'(\Delta)\le0
\quad\text{for all }\Delta\in(\underline{\Delta},0]\text{ and all }t,
\]
so that
\[
f_2'(\Delta)=\sum_t f_{2,t}'(\Delta)\le0,
\quad
\forall\,\Delta\in(\underline{\Delta},0].
\]
Hence $f_2(\Delta)$ is nonincreasing on $(\underline{\Delta},0]$, and the non-positive bias in the SW experiment follows from
\[
\EE\widehat{GTE}^{SW}-GTE = \frac{1}{NT}f_2(\Delta_1)-\frac{1}{NT}f_2(\Delta_0).
\]


\subsection{Proof of Theorem~\ref{thm:S1_IR}}

In the item-level randomized experiment, denote $W_{n,\cdot}^{IR} := W_{n,t}^{IR}$ for all $t \in [T]$. Then
\begin{align*}
    \EE \widehat{GT}^{IR} 
    & = \frac{1}{NT} \sum_{n=1}^N 
        \EE \Big[  W_{n,\cdot}^{IR}\frac{\sum_{t=1}^T  R_{n,t}}{p} \Big]\\
    & = \frac{1}{NT} \sum_{n=1}^N 
        \EE \Big[ W_{n,\cdot}^{IR} \frac{\sum_{t=1}^T   R_{n,t}^+(\hat S_{n,t},D_{n,t}) }{p} \Big]\\
    & =\frac{1}{NT}  \sum_{n=1}^N 
        \frac{\PP\big[ W_{n,\cdot}^{IR} = 1 \big]}{p} 
        \EE\Big[\sum_{t=1}^T  R_{n,t}^+(\hat S_{n,t},D_{n,t}) \mid W_{n,\cdot}^{IR} = 1\Big] \\
    & =  \frac{1}{NT} \sum_{n=1}^N  \sum_{t=1}^T 
        \EE \Big[ R_{n,t}^+(\hat S_{n,t}(\W_t^{IR}),D_{n,t}) \mid W_{n,t}^{IR}=1\Big],
\end{align*}
where the second equality uses \eqref{eq: sum_r_+}.

Combining with \eqref{eq: GT}, it holds that 
\begin{align}
\EE \widehat{GT}^{IR} - GT
= \frac{1}{NT}\sum_{n=1}^N \sum_{t=1}^T 
\Big(
    \EE[ R_{n,t}^+(\hat S_{n,t}(\W_t^{IR}),D_{n,t}) \mid W_{n,t}^{IR} = 1]
    - \EE R_{n,t}^+(\hat S_{n,t}(\boldsymbol{1}),D_{n,t})
\Big).
\label{eq:IR_T_bias}
\end{align}

Similarly, for the control group, it holds that
\begin{align}
\EE \widehat{GC}^{IR} - GC
= \frac{1}{NT}\sum_{n=1}^N \sum_{t=1}^T 
\Big(
    \EE[ R_{n,t}^+(\hat S_{n,t}(\W_t^{IR}),D_{n,t}) \mid W_{n,t}^{IR} = 0]
    - \EE R_{n,t}^+(\hat S_{n,t}(\boldsymbol{0}),D_{n,t})
\Big).
\label{eq:IR_C_bias}
\end{align}

Therefore,
\begin{align*}
     \EE \widehat{GTE}^{IR}  - GTE 
     & =
        \frac{1}{NT}\sum_{n=1}^N \sum_{t=1}^T
        \Big(
            \EE[ R_{n,t}^+(\hat S_{n,t}(\W_t^{IR}),D_{n,t}) \mid W_{n,t}^{IR} = 1]
            - \EE R_{n,t}^+(\hat S_{n,t}(\boldsymbol{1}),D_{n,t})
        \Big)   \\
     &\quad -
        \frac{1}{NT}\sum_{n=1}^N \sum_{t=1}^T
        \Big(
            \EE[ R_{n,t}^+(\hat S_{n,t}(\W_t^{IR}),D_{n,t}) \mid W_{n,t}^{IR} = 0]
            - \EE R_{n,t}^+(\hat S_{n,t}(\boldsymbol{0}),D_{n,t})
        \Big). 
\end{align*}

Suppose $\hat \mu_{n,t}(0) < \hat \mu_{n,t}(1) \leq \mu_{n,t}$, so that both forecasts underestimate demand but the treatment is less biased. Under Assumptions~\ref{asp:no_overshoot} and ~\ref{asp:margin_bounded_multiplier} , we have $I_{n,t} \leq \hat S_{n,t}=\bar S_{n,t}$ for any $t \in [T]$ and $n \in [N]$. Therefore, for each assignment $\boldsymbol{w} \in \{0,1\}^N$, the base-stock level is
\[
\hat S_{n,t}(\boldsymbol{w})
= \hat \mu_{n,t}(w_n) 
  + \Bigg(2\Bigg(\frac{m_n-\hat \lambda_t(\boldsymbol{w})}{M_n}\Bigg) - 1 \Bigg)\alpha_{n,t},
\]
where $\hat \lambda_t(\boldsymbol{w})$ is the shadow price associated with the capacity constraint under assignment $\boldsymbol{w}$.

Consider the function $\EE_{D_{n,t}} R_{n,t}^+(s,D_{n,t})$ with respect to $s$. We have
\[
\frac{d}{ds}\EE R_{n,t}^+(s,D_{n,t})
= m_n - M_n\,F_{n,t}(s),
\]
so $\EE R_{n,t}^+(s,D_{n,t})$ is increasing in $s$ as long as
\[
F_{n,t}(s) \le \frac{m_n}{M_n}.
\]
Let 
\[
q_{n,t}:=F_{n,t}^{-1}\Big(\frac{m_n}{M_n}\Big)
\]
be this threshold. We next argue that all relevant base-stock levels in the IR experiment lie below $q_{n,t}$.

For any assignment $\boldsymbol{w}$, let $\widehat F_{n,t}^{\boldsymbol{w}}$ denote the estimated CDF used for item $n$ in period $t$ under assignment $\boldsymbol{w}$, i.e.,
\[
\widehat D_{n,t}^{\boldsymbol{w}}
\sim \mathrm{Unif}\big[
    \hat\mu_{n,t}(w_n)-\alpha_{n,t},
    \hat\mu_{n,t}(w_n)+\alpha_{n,t}
\big].
\]
By the inherited margin bound under Assumption~\ref{asp:margin_bounded_multiplier}, the KKT condition for the myopic base-stock problem is written with respect to the estimated distribution:
\[
\widehat F_{n,t}^{\boldsymbol{w}}\big(\hat S_{n,t}(\boldsymbol{w})\big)
= \frac{m_n-\hat\lambda_t(\boldsymbol{w})}{M_n}
\le \frac{m_n}{M_n}.
\]
Since $\hat\mu_{n,t}(w_n)\le \mu_{n,t}$ and $\widehat\alpha_{n,t}=\alpha_{n,t}$, the estimated demand distribution is a leftward shift of the true demand distribution. Hence, for every $s\in\mathbb R$,
\[
F_{n,t}(s)\le \widehat F_{n,t}^{\boldsymbol{w}}(s).
\]
Applying this inequality at $s=\hat S_{n,t}(\boldsymbol{w})$ yields
\[
F_{n,t}\big(\hat S_{n,t}(\boldsymbol{w})\big)
\le 
\widehat F_{n,t}^{\boldsymbol{w}}\big(\hat S_{n,t}(\boldsymbol{w})\big)
\le 
\frac{m_n}{M_n}.
\]
Since $F_{n,t}$ is increasing, it follows that
\[
\hat S_{n,t}(\boldsymbol{w}) \le q_{n,t},
\]
for any assignment $\boldsymbol{w}$.

Now consider all $\boldsymbol{w}$ with $W_{n,t}=1$. Since $\hat\mu_{k,t}(0)<\hat\mu_{k,t}(1)$ for every item $k$, the estimated-mean vector under the global treatment assignment $\boldsymbol{1}$ is component-wise weakly larger than under any mixed assignment $\boldsymbol{w}$ with $W_{n,t}=1$. By the monotonicity of the Lagrange multiplier with respect to the demand parameters,
\[
    \hat\lambda_t(\boldsymbol{w}) \le \hat\lambda_t(\boldsymbol{1}).
\]
For item $n$, the own estimated mean is the same under $\boldsymbol{w}$ and $\boldsymbol{1}$ because $W_{n,t}=1$. Therefore, 
\begin{align}
        \hat S_{n,t}(\boldsymbol{1}) \le \hat S_{n,t}(\boldsymbol{w}) \le q_{n,t}.
\label{eq:basestock_treat_vs_GT}
\end{align}

Similarly, for all $\boldsymbol{w}$ with $W_{n,t}=0$, the estimated-mean vector under $\boldsymbol{w}$ is component-wise weakly larger than under the global control assignment $\boldsymbol{0}$. Hence,
\[
    \hat\lambda_t(\boldsymbol{0}) \le \hat\lambda_t(\boldsymbol{w}).
\]
For item $n$, the own estimated mean is the same under $\boldsymbol{w}$ and $\boldsymbol{0}$ because $W_{n,t}=0$. Therefore,
\begin{align}
    \hat S_{n,t}(\boldsymbol{w}) \le \hat S_{n,t}(\boldsymbol{0}) \le q_{n,t}.
\label{eq:basestock_control_vs_GC}
\end{align}

Since $\EE R_{n,t}^+(s,D_{n,t})$ is increasing in $s$ on $(-\infty,q_{n,t}]$, we obtain
\begin{align}
    \EE[ R_{n,t}^+( \hat S_{n,t}(\W_t^{IR}),D_{n,t}) \mid W_{n,t}^{IR} = 1] 
    \geq  \EE R_{n,t}^+( \hat S_{n,t}(\boldsymbol{1}),D_{n,t}),
    \label{eq: IR_treatment}
\end{align}
and
\begin{align}
       \EE[ R_{n,t}^+( \hat S_{n,t}(\W_t^{IR}),D_{n,t}) \mid W_{n,t}^{IR} = 0] 
       \leq  \EE R_{n,t}^+( \hat S_{n,t}(\boldsymbol{0}),D_{n,t}).
       \label{eq: IR_control}
\end{align}
Combining \eqref{eq:IR_T_bias}, \eqref{eq:IR_C_bias}, \eqref{eq: IR_treatment}, and \eqref{eq: IR_control}, we obtain the non-negativity of $Bias^{IR}$.

\subsection{Proof of Theorem~\ref{thm:S1_PR}}

In the pairwise randomized experiment, 
\begin{align}
     &\quad  \sum_{t=1}^T   \EE \big[ R_{n,t} \mid W_{n,t}^{PR} = 1\big] \notag \\
     & =  \sum_{t=1}^T\EE\Big[ R^+_{n,t}(\hat S_{n,t}(\W_t^{PR}),D_{n,t})  + c_n  I_{n,t} - c_n (\hat S_{n,t}(\W_t^{PR}) - D_{n,t})^+ \mid W_{n,t}^{PR} = 1 \Big] \notag \\
     &\quad + c_n \EE\Big[(\hat S_{n,T}(\W_T^{PR}) - D_{n,T})^+ \mid W_{n,T}^{PR} = 1\Big] \notag \\
     &= \sum_{t=1}^T\EE \Big[ R^+_{n,t}(\hat S_{n,t}(\W_t^{PR}),D_{n,t})\mid W_{n,t}^{PR} = 1\Big] + \sum_{t=1}^T c_n\EE \Big[I_{n,t} \mid W_{n,t}^{PR} = 1\Big] \notag\\
     &\quad - \sum_{t=1}^{T-1} c_n \EE \Big[(\hat S_{n,t}(\W_t^{PR}) - D_{n,t})^+ \mid W_{n,t}^{PR} = 1\Big].   \label{eq: PR1}
\end{align}
For $t \geqslant 2$, we have
\begin{align}
    \quad & \EE \Big[I_{n,t} \mid W_{n,t}^{PR} = 1\Big] \nonumber\\
    &  = \EE\Big[(\hat S_{n,t-1}(\W_{t-1}^{PR}) - D_{n,t-1})^+\mid W_{n,t}^{PR} = 1\Big] \notag \\
    &= p \EE  \Big[(\hat S_{n,t-1}(\W_{t-1}^{PR}) - D_{n,t-1})^+ \mid W_{n,t-1}^{PR} = 1\Big]   + (1-p) \EE \Big[(\hat S_{n,t-1}(\W_{t-1}^{PR}) - D_{n,t-1})^+ \mid W_{n,t-1}^{PR} = 0\Big],  \label{eq: PR2}
\end{align}
where the last equality is due to $\W_{t-1}^{PR} \perp W_{n,t}^{PR}$.
Substituting \eqref{eq: PR2} into \eqref{eq: PR1}, we obtain
\begin{align*}
  & \quad \sum_{t=1}^T   \EE \Big[ R_{n,t} \mid W_{n,t}^{PR} = 1\Big] \\
  &=  \sum_{t=1}^T\EE\Big[ R^+_{n,t}(\hat S_{n,t}(\W_t^{PR}),D_{n,t})\mid W_{n,t}^{PR} = 1\Big] + c_n I_{n,1} \\
 &\quad - \sum_{t=1}^{T-1} c_n(1-p) \Big( \EE \big[(\hat S_{n,t}(\W_{t}^{PR}) - D_{n,t})^+ \mid W_{n,t}^{PR} = 1\big]  -  \EE \big[(\hat S_{n,t}(\W_{t}^{PR}) - D_{n,t})^+ \mid W_{n,t}^{PR} = 0\big] \Big).
\end{align*}

Note that $I_{n,1} = 0$ for all $n$. Combining with \eqref{eq: GT}, it holds that 
\begin{align*}
  &  \quad   \EE \widehat{GT}^{PR} - GT\\
   & =  \frac{1}{NT}\sum_{n=1}^N \sum_{t=1}^T\Big(\EE[ R_{n,t}^+(\hat S_{n,t}(\W_t^{PR}),D_{n,t}) \mid W_{n,t}^{PR} = 1] - \EE R_{n,t}^+(\hat S_{n,t}(\boldsymbol{1}),D_{n,t}) \Big)  \\
 &\quad - \frac{1}{NT}\sum_{n=1}^N\sum_{t=1}^{T-1} c_n(1-p) \Big( \EE \big[(\hat S_{n,t}(\W_{t}^{PR}) - D_{n,t})^+ \mid W_{n,t}^{PR} = 1\big]  -  \EE \big[(\hat S_{n,t}(\W_{t}^{PR}) - D_{n,t})^+ \mid W_{n,t}^{PR} = 0\big] \Big).
\end{align*}

Similarly, for the control group, 
\begin{align*}
    \EE \widehat{GC}^{PR} - GC & =             \frac{1}{NT}\sum_{n=1}^N \sum_{t=1}^T\Big(\EE[ R_{n,t}^+(\hat S_{n,t}(\W_t^{PR}),D_{n,t}) \mid W_{n,t}^{PR} = 0] - \EE R_{n,t}^+(\hat S_{n,t}(\boldsymbol{0}),D_{n,t}) \Big)  \\
 &\quad + \frac{1}{NT}\sum_{n=1}^N \sum_{t=1}^{T-1} c_np \Big( \EE \big[(\hat S_{n,t}(\W_{t}^{PR}) - D_{n,t})^+ \mid W_{n,t}^{PR} = 1\big]  -  \EE \big[(\hat S_{n,t}(\W_{t}^{PR}) - D_{n,t})^+ \mid W_{n,t}^{PR} = 0\big] \Big).
\end{align*}

Therefore, we obtain
\begin{align*}
     \EE \widehat{GTE}^{PR}  - GTE & =
            \frac{1}{NT}\sum_{n=1}^N \sum_{t=1}^T\Big(\EE[ R_{n,t}^+(\hat S_{n,t}(\W_t^{PR}),D_{n,t}) \mid W_{n,t}^{PR} = 1] - \EE R_{n,t}^+(\hat S_{n,t}(\boldsymbol{1}),D_{n,t}) \Big)   \\
       &\quad - \frac{1}{NT}\sum_{n=1}^N \sum_{t=1}^T\Big(\EE[ R_{n,t}^+(\hat S_{n,t}(\W_t^{PR}),D_{n,t}) \mid W_{n,t}^{PR} = 0] - \EE R_{n,t}^+(\hat S_{n,t}(\boldsymbol{0}),D_{n,t}) \Big)   \\
   &\quad -  \frac{1}{NT} \sum_{n=1}^N \sum_{t=1}^{T-1} c_n \Big(\EE[(\hat S_{n,t}(\W_t^{PR}) - D_{n,t})^+ \mid W_{n,t}^{PR} = 1]  - \EE[(\hat S_{n,t}(\W_t^{PR}) - D_{n,t})^+ \mid W_{n,t}^{PR} = 0] \Big). 
\end{align*}

Notice that for a fixed $t$, the distribution of $[\W_t \mid W_{n,t} = 1]$ is the same in IR and PR experimental designs for all $t\in [T]$. Combining with the bias of the IR estimator, we have
\begin{align*}
    & \quad \EE \widehat{GTE}^{PR}  - \EE \widehat{GTE}^{IR}\\  
    & = - \frac{1}{NT} \sum_{n=1}^N \sum_{t=1}^{T-1} c_n \Big(\EE[(\hat S_{n,t}(\W_t^{PR}) - D_{n,t})^+ \mid W_{n,t}^{PR} = 1]  - \EE[(\hat S_{n,t}(\W_t^{PR}) - D_{n,t})^+ \mid W_{n,t}^{PR} = 0] \Big).
\end{align*}

\vspace{10pt}

Fix an item $n$ and a period $t$. Fix arbitrarily the treatment assignments of
all other items at time $t$, i.e., fix $W_{m,t}^{PR}$ for all $m\neq n$. Let
$\bw_t^{(1)}$ and $\bw_t^{(0)}$ denote the two assignment vectors that coincide
on all coordinates $m\neq n$ and differ only in the $n$-th coordinate:
$W_{n,t}^{PR}=1$ in $\bw_t^{(1)}$ and $W_{n,t}^{PR}=0$ in $\bw_t^{(0)}$. Under these two
assignments, the estimated means for item $n$ satisfy
\[
\hat\mu_{n,t}(0)<\hat\mu_{n,t}(1).
\]

For this fixed configuration of other items, consider the base-stock level for
item $n$ as a function of its own estimated mean. With a slight abuse of notation, let
$s\mapsto \hat S_{n,t}(s)$ be the solution of the KKT system when we replace
$\hat\mu_{n,t}$ by a generic value $s$ and keep all other inputs (including
$\hat\mu_{m,t}(W_{m,t}^{PR})$ for $m\neq n$) fixed. Then
\[
\hat S_{n,t}(\bw_t^{(1)}) = \hat S_{n,t}(\hat\mu_{n,t}(1)),
\qquad
\hat S_{n,t}(\bw_t^{(0)}) = \hat S_{n,t}(\hat\mu_{n,t}(0)).
\]

Under Assumptions~\ref{asp:no_overshoot} and ~\ref{asp:margin_bounded_multiplier} , we have $I_{n,t}\leq \hat {S}_{n,t}=\bar S_{n,t}$, and the base-stock level is given by
\[
\hat S_{k,t}(\W_t)
= \hat \mu_{k,t}(W_{k,t}) 
  + \Bigl(2 \Bigl(\tfrac{m_k-\hat\lambda_t(\W_t)}{M_k}\Bigr) - 1 \Bigr)\alpha_{k,t}, \quad \forall k \in [N].
\]
We compare the assignments $\bw_t^{(1)}$  and $\bw_t^{(0)}$, where $w_{n,t}^{(1)}=1$, $w_{n,t}^{(0)}=0$ and $w_{m,t}^{(1)} = w_{m,t}^{(0)}$ for all $m\neq n$. Since $\hat \mu_{n,t}(1) > \hat \mu_{n,t}(0)$ and the estimated means for all other items $m \ne n$ remain unchanged, the aggregate demand pressure increases. Consequently, the shadow price is non-decreasing:
\[
\hat \lambda_t(\bw_t^{(1)}) \ge \hat \lambda_t(\bw_t^{(0)}).
\]
For any other item $m \neq n$, the mean $\hat \mu_{m,t}$ is constant. Since the function $x \mapsto (x)^+$ is non-decreasing, the increase in $\hat \lambda_t$ implies
\[
\hat S_{m,t}(\bw_t^{(1)}) \le \hat S_{m,t}(\bw_t^{(0)}), \quad \forall m \neq n.
\]
To establish the inequality for item $n$, we consider the capacity constraint. There are two cases:

\begin{itemize}
    \item \textbf{Case 1:} The capacity constraint is binding under assignment $\bw_t^{(1)}$, i.e., $\sum_{k=1}^N \hat S_{k,t}(\bw_t^{(1)}) = B$.
    Since the total allocation under $\bw_t^{(0)}$ cannot exceed capacity ($\sum_{k=1}^N \hat S_{k,t}(\bw_t^{(0)}) \le B$), we have
    \[
    \sum_{k=1}^N \hat S_{k,t}(\bw_t^{(1)}) \ge \sum_{k=1}^N \hat S_{k,t}(\bw_t^{(0)}).
    \]
    Subtracting the sum over $m \neq n$ from both sides, and using the fact that $\sum_{m \neq n} \hat S_{m,t}(\bw_t^{(1)}) \le \sum_{m \neq n} \hat S_{m,t}(\bw_t^{(0)})$, we obtain
    \[
    \hat S_{n,t}(\bw_t^{(1)}) \ge \hat S_{n,t}(\bw_t^{(0)}).
    \]

    \item \textbf{Case 2:} The capacity constraint is slack under assignment $\bw_t^{(1)}$, i.e., $\sum_{k=1}^N \hat S_{k,t}(\bw_t^{(1)}) < B$.
    In this case, the shadow price must be zero: $\hat \lambda_t(\bw_t^{(1)}) = 0$. Since Lagrange multipliers are non-negative and non-decreasing in demand, this implies $\hat \lambda_t(\bw_t^{(0)}) = 0$ as well.
    With $\hat \lambda_t = 0$ in both cases, the base-stock level is determined purely by the mean. Since $\hat \mu_{n,t}(1) > \hat \mu_{n,t}(0)$, it follows directly that
    \[
    \hat S_{n,t}(\bw_t^{(1)}) > \hat S_{n,t}(\bw_t^{(0)}).
    \]
\end{itemize}
Combining both cases, we conclude that
\[
\hat S_{n,t}(\bw_t^{(1)}) \ge \hat S_{n,t}(\bw_t^{(0)}).
\]

For any realization of $D_{n,t}$, the function $s\mapsto (s-D_{n,t})^+$ is nondecreasing, so for every fixed configuration of
$(W_{m,t}^{PR})_{m\neq n}$ and for every outcome of $D_{n,t}$ we have
\[
\bigl(\hat S_{n,t}(\bw_t^{(1)})-D_{n,t}\bigr)^+
\;\ge\;
\bigl(\hat S_{n,t}(\bw_t^{(0)})-D_{n,t}\bigr)^+.
\]
Taking expectations over $(\W_t^{PR},D_{n,t})$ and conditioning on $W_{n,t}^{PR}$ yields
\[
\EE\bigl[(\hat S_{n,t}(\W_t^{PR})-D_{n,t})^+ \mid W_{n,t}^{PR}=1\bigr]
-
\EE\bigl[(\hat S_{n,t}(\W_t^{PR})-D_{n,t})^+ \mid W_{n,t}^{PR}=0\bigr]
\;\ge\; 0,
\]
and therefore
\begin{align*}
   &\quad  \EE \widehat{GTE}^{PR}  - \EE \widehat{GTE}^{IR}  \\
&     = - \frac{1}{NT} \sum_{n=1}^N \sum_{t=1}^{T-1} c_n \Big(\EE[(\hat S_{n,t}(\W_t^{PR}) - D_{n,t})^+ \mid W_{n,t}^{PR} = 1]  - \EE[(\hat S_{n,t}(\W_t^{PR}) - D_{n,t})^+ \mid W_{n,t}^{PR} = 0] \Big) \leq 0.
\end{align*}
This implies $Bias^{PR} \le Bias^{IR}$.

\section{Proofs for Section~\ref{section:scenario2}}
\label{appendix:scenario2-proofs}

This appendix proves the Scenario~2 mean-field results. We first give  conditions that justify the affine-response region used in Assumption~\ref{asp:s2_margin}. We then prove convergence of the random KKT multiplier and use that convergence to establish the asymptotic signs of GTE, SW bias, IR bias, and PR bias under Assumption~\ref{asp:no_overshoot} and  Assumptions~\ref{asp:s2_margin}--\ref{asp:unique_zero}.

This appendix provides the proofs for Lemma~\ref{lem:lambda_limit}, 
Proposition~\ref{thm:S2_gte}  and Theorems~\ref{thm:S2_sw}–\ref{thm:S2_pr} in Section~\ref{section:scenario2}.
Throughout, we work under Assumption~\ref{asp:no_overshoot} and  Assumptions~\ref{asp:s2_margin}--\ref{asp:unique_zero}.

\subsection{Verification of Scenario~2 affine-response validity}
\label{app:s2-margin-verification}
The goal is to choose an error radius so that every possible forecast realization keeps the multiplier interior, keeps estimated lower support nonnegative, and prevents inventory overshoot across periods as required in Assumption~\ref{asp:no_overshoot} and Assumption~\ref{asp:s2_margin}. 

For each $N$ and $t$, let
\[
A_t^{(N)}:=\frac{2}{N}\sum_{n=1}^N \alpha_{n,t}v_n>0,
\qquad
C_t^{(N)}
:=
\frac1N\sum_{n=1}^N
\Bigl\{\mu_{n,t}+\alpha_{n,t}(2u_n-1)\Bigr\}
-\frac{B^{(N)}}{N},
\]
\[
\overline\lambda_t^{(N)}(\eta)
:=
\left(\frac{C_t^{(N)}+\eta}{A_t^{(N)}}\right)^+,
\qquad
m_{\min}^{(N)}:=\min_{n\in[N]}m_n,
\]
and
\[
\overline q_{n,t}^{(N)}(\eta)
:=
\mu_{n,t}+\eta+\alpha_{n,t}(2u_n-1),
\]
\[
\underline q_{n,t}^{(N)}(\eta)
:=
\mu_{n,t}-\eta+
\alpha_{n,t}
\Bigl(2u_n-1-2v_n\overline\lambda_t^{(N)}(\eta)\Bigr).
\]
The following three inequalities are sufficient:
\begin{align}
\overline\lambda_t^{(N)}(\eta)<m_{\min}^{(N)},
\qquad \forall N,t,
\label{eq:eta_interior_lambda_bound}
\end{align}
\begin{align}
\overline q_{n,t}^{(N)}(\eta)
-
\underline q_{n,t+1}^{(N)}(\eta)
\le
\mu_{n,t}-\alpha_{n,t},
\qquad
\forall N,\ n\in[N],\ t\in[T-1],
\label{eq:eta_no_overshoot_bound}
\end{align}
and
\begin{align}
\mu_{n,t}-\alpha_{n,t}-\eta \ge 0,
\qquad \forall N,\ n\in[N],\ t\in[T].
\label{eq:eta_nonnegative_lower_support}
\end{align}
Condition~\eqref{eq:eta_interior_lambda_bound} is equivalent to $(C_t^{(N)}+\eta)^+<A_t^{(N)}m_{\min}^{(N)}$.

\begin{lemma}[Scenario~2 affine-response verification]
\label{lem:s2_margin_verification}
If \eqref{eq:eta_interior_lambda_bound}, \eqref{eq:eta_no_overshoot_bound}, and \eqref{eq:eta_nonnegative_lower_support} hold, then, for every $N$, period $t$, assignment vector $\W_t$, and forecast-error realization satisfying $|\epsilon_{n,t}(w)|\le\eta$, the predictive KKT multiplier satisfies
\[
0\le \hat\lambda_t^{(N)}(\W_t)
\le \overline\lambda_t^{(N)}(\eta)
< m_{\min}^{(N)}.
\]
Moreover, the estimated lower support is nonnegative and the no-overshoot condition holds uniformly over this bounded-error class. Consequently, the affine-response validity part of Assumption~\ref{asp:s2_margin} follows from these three envelope conditions.
\end{lemma}

\begin{proof}
Fix $N$, $t$, an assignment vector $\W_t$, and a realization satisfying $|\epsilon_{n,t}(w)|\le\eta$. Consider the candidate expression
\[
\widetilde S_{n,t}^{(N)}(\lambda)
:=
\mu_{n,t}+\epsilon_{n,t}(W_{n,t})
+\alpha_{n,t}(2u_n-1-2v_n\lambda),
\]
and the corresponding aggregate slack
\[
\widetilde\Psi_t^{(N)}(\lambda)
:=
\frac1N\sum_{n=1}^N \widetilde S_{n,t}^{(N)}(\lambda)
-\frac{B^{(N)}}{N}.
\]
Let
\[
\widetilde\lambda_t^{(N)}(\W_t)
:=
\max\{0,\inf\{\lambda\ge0:\widetilde\Psi_t^{(N)}(\lambda)\le0\}\}.
\]
For any $\lambda\ge0$,
\begin{align*}
\widetilde\Psi_t^{(N)}(\lambda)
&=
\frac1N\sum_{n=1}^N
\Bigl\{
\mu_{n,t}+\epsilon_{n,t}(W_{n,t})
+\alpha_{n,t}(2u_n-1-2v_n\lambda)
\Bigr\}
-\frac{B^{(N)}}{N} \\
&\le
C_t^{(N)}+\eta-A_t^{(N)}\lambda.
\end{align*}
At $\lambda=\overline\lambda_t^{(N)}(\eta)$, the right-hand side is nonpositive. Since $\widetilde\Psi_t^{(N)}(\lambda)$ is nonincreasing in $\lambda$,
\[
0\le \widetilde\lambda_t^{(N)}(\W_t)
\le \overline\lambda_t^{(N)}(\eta).
\]
By \eqref{eq:eta_interior_lambda_bound},
\[
\widetilde\lambda_t^{(N)}(\W_t)<m_{\min}^{(N)}.
\]
Thus every item's critical ratio is strictly positive under this multiplier.

The envelope condition verifies no overshoot. For any bounded-error realization and any assignment, the largest possible value of
$\widetilde S_{n,t}^{(N)}(\widetilde\lambda_t^{(N)}(\W_t))$ is bounded above by $\overline q_{n,t}^{(N)}(\eta)$, because the forecast error is at most $\eta$ and the multiplier is nonnegative. Similarly, the smallest possible value of
$\widetilde S_{n,t+1}^{(N)}(\widetilde\lambda_{t+1}^{(N)}(\W_{t+1}))$ is bounded below by $\underline q_{n,t+1}^{(N)}(\eta)$, because the forecast error is at least $-\eta$ and the multiplier is at most $\overline\lambda_{t+1}^{(N)}(\eta)$. Hence \eqref{eq:eta_no_overshoot_bound} implies
\[
\widetilde S_{n,t}^{(N)}(\widetilde\lambda_t^{(N)}(\W_t))
-
\widetilde S_{n,t+1}^{(N)}(\widetilde\lambda_{t+1}^{(N)}(\W_{t+1}))
\le
\mu_{n,t}-\alpha_{n,t},
\]
which is the no-overshoot condition.

The strict margin bound makes the positive-part truncation in the newsvendor critical ratio inactive. In addition, \eqref{eq:eta_nonnegative_lower_support} gives \(\hat\mu_{n,t}(W_{n,t})-\alpha_{n,t}\ge\mu_{n,t}-\alpha_{n,t}-\eta\ge0\), so the estimated lower-support condition holds. The same inequality also implies
\[
\widetilde S_{n,1}^{(N)}(\widetilde\lambda_1^{(N)}(\W_1))\ge 0=I_{n,1},
\]
because $\epsilon_{n,1}(W_{n,1})\ge-\eta$ and $u_n-v_n\widetilde\lambda_1^{(N)}(\W_1)>0$. Suppose now that
$I_{n,t}\le \widetilde S_{n,t}^{(N)}(\widetilde\lambda_t^{(N)}(\W_t))$. Then
\[
I_{n,t+1}
=
\left(\widetilde S_{n,t}^{(N)}(\widetilde\lambda_t^{(N)}(\W_t))-D_{n,t}\right)^+
\le
\widetilde S_{n,t+1}^{(N)}(\widetilde\lambda_{t+1}^{(N)}(\W_{t+1})),
\]
where the last inequality follows from the no-overshoot condition and the nonnegativity of the next-period target. Hence the inventory lower bound does not bind in any period. Therefore the exact predictive KKT solution coincides with the candidate solution:
\[
\hat\lambda_t^{(N)}(\W_t)=\widetilde\lambda_t^{(N)}(\W_t),
\qquad
\hat S_{n,t}^{(N)}(\W_t)=
\widetilde S_{n,t}^{(N)}(\widetilde\lambda_t^{(N)}(\W_t)).
\]
Consequently,
\[
0\le \hat\lambda_t^{(N)}(\W_t)
\le \overline\lambda_t^{(N)}(\eta)
< m_{\min}^{(N)},
\]
and the no-overshoot condition holds for the actual predictive base-stock levels.
\end{proof}

\begin{remark}[Non-vacuity of the Scenario~2 affine-response condition]
The conditions above are compatible whenever the corresponding zero-error system has strict slack. In particular, it is sufficient that
\[
(C_t^{(N)})^+<A_t^{(N)}m_{\min}^{(N)},\qquad
\mu_{n,t}-\alpha_{n,t}>0,
\]
and
\[
\overline q_{n,t}^{(N)}(0)-\underline q_{n,t+1}^{(N)}(0)
<
\mu_{n,t}-\alpha_{n,t}
\]
hold uniformly over the relevant $N,n,t$. Under these strict inequalities, continuity in $\eta$ implies that there exists a sufficiently small positive radius $\eta$ satisfying
\eqref{eq:eta_interior_lambda_bound}, \eqref{eq:eta_no_overshoot_bound}, and
\eqref{eq:eta_nonnegative_lower_support}.
\end{remark}

\subsection{Proof of Lemma~\ref{lem:lambda_limit}}

We work on a probability space $(\Omega,\mathcal F,\PP)$ supporting all treatment assignments and forecast errors. Fix a period $t$ and suppress $t$ when there is no ambiguity. We first prove the claim for a generic assignment vector independent of the forecast errors; deterministic assignment vectors are covered by the same argument by treating the assignments as constants.

\emph{Step 1: Law of large numbers for the forecast component.}
Write
\[
\epsilon_n:=\hat\mu_n(W_n)-\mu_n=\epsilon_n(W_n).
\]
Let $\mathcal F^W:=\sigma(\{W_m:m\in\mathbb N\})$ be the $\sigma$-field generated by the assignments. Conditional on $\mathcal F^W$, the random variables $\epsilon_n(W_n)$ are independent across $n$, centered, and uniformly bounded by $\eta$ under Assumptions~\ref{asp:s2_margin}, and \ref{asp:independence}. Hence Kolmogorov's strong law for independent uniformly bounded arrays yields
\[
\frac1N\sum_{n=1}^N \epsilon_n(W_n)\xrightarrow[N\to\infty]{a.s.}0.
\]
Combining this with Assumption~\ref{asp:para_bounded_limit}, we obtain
\[
\frac1N\sum_{n=1}^N \hat\mu_n(W_n)\xrightarrow[N\to\infty]{a.s.}\mu_0.
\]

\emph{Step 2: Pointwise convergence of the candidate aggregate slack.}
For $\lambda\in[0,m_*)$, define
\[
\widetilde S_n^{(N)}(\lambda)
:=\hat\mu_n(W_n)+\alpha_n\Bigl(2(u_n-v_n\lambda)-1\Bigr),
\]
and
\[
\widetilde\Psi_t^{(N)}(\lambda)
:=\frac1N\sum_{n=1}^N \widetilde S_n^{(N)}(\lambda)-\frac{B^{(N)}}{N}.
\]
For each fixed $\lambda\in[0,m_*)$,
\begin{align*}
\widetilde\Psi_t^{(N)}(\lambda)
&=\frac1N\sum_{n=1}^N \hat\mu_n(W_n)
+\frac1N\sum_{n=1}^N \alpha_n\Bigl(2(u_n-v_n\lambda)-1\Bigr)
-\frac{B^{(N)}}{N} \\
&\xrightarrow[N\to\infty]{a.s.}
\psi_t(\lambda):=\mu_0+k_t(\lambda)-\beta_0,
\end{align*}
where the convergence of the first term follows from Step~1, and the convergence of the second and third terms follows from Assumptions~\ref{asp:para_bounded_limit}--\ref{asp:kt_limit}.

\emph{Step 3: Convergence of the candidate multiplier.}
Define
\[
\widetilde\lambda_t^{(N)}(\W_t)
:=\max\{0,\inf\{\lambda\ge0:\widetilde\Psi_t^{(N)}(\lambda)\le0\}\}.
\]
The function $\lambda\mapsto\widetilde\Psi_t^{(N)}(\lambda)$ is nonincreasing. We prove convergence separately for the binding and slack mean-field cases in Assumption~\ref{asp:unique_zero}.

First suppose $\psi_t(0)>0$, so $\lambda_t^*>0$. Pick $\lambda^-<\lambda_t^*<\lambda^+$, both in $[0,m_*)$, such that $\psi_t(\lambda^-)>0$ and $\psi_t(\lambda^+)<0$. By Step~2, almost surely, for all sufficiently large $N$,
\[
\widetilde\Psi_t^{(N)}(\lambda^-)>0,
\qquad
\widetilde\Psi_t^{(N)}(\lambda^+)<0.
\]
Since $\widetilde\Psi_t^{(N)}$ is nonincreasing, the smallest candidate multiplier satisfying the capacity inequality lies in $[\lambda^-,\lambda^+]$ for all sufficiently large $N$. Letting $\lambda^-\uparrow\lambda_t^*$ and $\lambda^+\downarrow\lambda_t^*$ along deterministic sequences gives
\[
\widetilde\lambda_t^{(N)}(\W_t)\xrightarrow[N\to\infty]{a.s.}\lambda_t^*.
\]

Now suppose $\psi_t(0)\le0$, so $\lambda_t^*=0$ by Assumption~\ref{asp:unique_zero}. For any $\lambda^+\in(0,m_*)$, Assumption~\ref{asp:unique_zero} gives $\psi_t(\lambda^+)<0$. By Step~2, $\widetilde\Psi_t^{(N)}(\lambda^+)<0$ eventually almost surely, and hence $0\le \widetilde\lambda_t^{(N)}(\W_t)\le\lambda^+$ eventually. Letting $\lambda^+\downarrow0$ yields
\[
\widetilde\lambda_t^{(N)}(\W_t)\xrightarrow[N\to\infty]{a.s.}0=\lambda_t^*.
\]
Thus the candidate multiplier converges almost surely to $\lambda_t^*$ in both cases.

\emph{Step 4: Identification with the exact predictive multiplier and $L^1$ convergence.}
By Lemma~\ref{lem:s2_margin_verification}, the exact predictive KKT multiplier equals the candidate multiplier:
\[
\hat\lambda_t^{(N)}(\W_t)=\widetilde\lambda_t^{(N)}(\W_t).
\]
Therefore
\[
\hat\lambda_t^{(N)}(\W_t)\xrightarrow[N\to\infty]{a.s.}\lambda_t^*.
\]
Moreover, Lemma~\ref{lem:s2_margin_verification} and Assumption~\ref{asp:moment_bounds} give the deterministic bound
\[
0\le \hat\lambda_t^{(N)}(\W_t)
\le \overline\lambda_t^{(N)}(\eta)
< m_{\min}^{(N)}
\le \sup_nm_n<\infty.
\]
Dominated convergence then implies
\begin{align}
    \EE |\hat\lambda_t^{(N)}(\W_t)-\lambda_t^*| \xrightarrow[N\to\infty]{}0. \label{eq:lambda_L1_converge}
\end{align}

\emph{Step 5: Convergence of base-stock levels.}
By Lemma~\ref{lem:s2_margin_verification}, the exact predictive base-stock level satisfies
\begin{align*}
\hat S_{n,t}^{(N)}(\W_t)
&=\hat\mu_{n,t}(W_{n,t})+
\alpha_{n,t}\Bigl(2(u_n-v_n\hat\lambda_t^{(N)}(\W_t))-1\Bigr)\\
&\xrightarrow[N\to\infty]{a.s.}
\hat\mu_{n,t}(W_{n,t})+
\alpha_{n,t}\Bigl(2(u_n-v_n\lambda_t^*)-1\Bigr)
=:X_{n,t}(W_{n,t}).
\end{align*}
If $\mathbf W_t$ is replaced by a deterministic assignment vector independent of the forecast errors, the same proof applies: Step~1 becomes the strong law for independent centered bounded errors with deterministic choices of $w_n$, and the remaining steps are unchanged. In particular, the conclusions hold for the global assignments $\boldsymbol{1}$ and $\boldsymbol{0}$.
This completes the proof.

\subsection{Proof of Proposition~\ref{thm:S2_gte}}

Recall that
\[
GTE^{(N)} 
= \frac{1}{NT}\sum_{n=1}^N\sum_{t=1}^T
\Big(\EE R_{n,t}^+\big(\hat S_{n,t}^{(N)}(\boldsymbol{1}),D_{n,t}\big)
 -\EE R_{n,t}^+\big(\hat S_{n,t}^{(N)}(\boldsymbol{0}),D_{n,t}\big)\Big).
\]
Fix $t$ and $w\in\{0,1\}$. By Lemma~\ref{lem:lambda_limit}, applied to the deterministic global assignment $w\boldsymbol 1$,
\[
\hat\lambda_t^{(N)}(w\boldsymbol 1)\to \lambda_t^* \quad\text{in }L^1.
\]
Under Assumption~\ref{asp:no_overshoot} and Assumption~\ref{asp:s2_margin}, the affine representation is valid. Thus, for each $n\le N$,
\[
\hat S_{n,t}^{(N)}(w\boldsymbol 1)
=\hat\mu_{n,t}(w)+\alpha_{n,t}\Bigl(2(u_n-v_n\hat\lambda_t^{(N)}(w\boldsymbol 1))-1\Bigr),
\]
whereas
\[
X_{n,t}(w)=\hat\mu_{n,t}(w)+\alpha_{n,t}\Bigl(2(u_n-v_n\lambda_t^*)-1\Bigr).
\]
By Assumption~\ref{asp:moment_bounds}, there are finite constants $L_S$ and $L_R$, independent of $n,t,N$, such that
\[
\big|\hat S_{n,t}^{(N)}(w\boldsymbol 1)-X_{n,t}(w)\big|
\le L_S\big|\hat\lambda_t^{(N)}(w\boldsymbol 1)-\lambda_t^*\big|
\]
and
\[
\big|R_{n,t}^+(s,D_{n,t})-R_{n,t}^+(s',D_{n,t})\big|
\le L_R |s-s'|.
\]
Consequently,
\begin{align}
\frac1N\sum_{n=1}^N
\left|
\EE R_{n,t}^+\big(\hat S_{n,t}^{(N)}(w\boldsymbol 1),D_{n,t}\big)
-
\EE R_{n,t}^+\big(X_{n,t}(w),D_{n,t}\big)
\right|
\le L_RL_S\EE\big|\hat\lambda_t^{(N)}(w\boldsymbol 1)-\lambda_t^*\big|\to 0.
\label{eq:s2_global_revenue_average_convergence}
\end{align}

We now pass to the limiting base-stock levels before applying convex order. This step is important: at finite $N$, the base-stock level depends on the empirical multiplier $\hat\lambda_t^{(N)}$, which is capacity-coupled with the forecast errors, so the following convex-order comparison is not a finite-$N$ sign argument. For the limiting response, write
\[
X_{n,t}(w)=a_{n,t}+\epsilon_{n,t}(w),
\qquad
 a_{n,t}:=\mu_{n,t}+\alpha_{n,t}\Bigl(2(u_n-v_n\lambda_t^*)-1\Bigr).
\]
For each fixed $n,t$, the forecast errors are centered and satisfy
$\epsilon_{n,t}(1)\le_{\mathrm{cx}}\epsilon_{n,t}(0)$. Moreover, Assumption~\ref{asp:independence} ensures that $\epsilon_{n,t}(w)$ is independent of $D_{n,t}-\mu_{n,t}$. Since $s\mapsto (s-D_{n,t})^+$ is convex,
\[
\EE(X_{n,t}(1)-D_{n,t})^+
\le
\EE(X_{n,t}(0)-D_{n,t})^+.
\]
The centeredness of the errors gives $\EE X_{n,t}(1)=\EE X_{n,t}(0)$. Therefore,
\[
\phi_{n,t}^0
:=\EE R_{n,t}^+(X_{n,t}(1),D_{n,t})
-\EE R_{n,t}^+(X_{n,t}(0),D_{n,t})
\ge0.
\]
Define
\[
\phi_{n,t}^{(N)}:=
\EE R_{n,t}^+\big(\hat S_{n,t}^{(N)}(\boldsymbol 1),D_{n,t}\big)
-
\EE R_{n,t}^+\big(\hat S_{n,t}^{(N)}(\boldsymbol 0),D_{n,t}\big).
\]
Equation~\eqref{eq:s2_global_revenue_average_convergence}, applied with $w=1$ and $w=0$, implies
\[
\frac1N\sum_{n=1}^N |\phi_{n,t}^{(N)}-\phi_{n,t}^0|\to 0
\quad\text{for each fixed }t.
\]
Since $T$ is fixed,
\begin{align*}
\liminf_{N\to\infty}GTE^{(N)}
&=\liminf_{N\to\infty}\frac1T\sum_{t=1}^T\frac1N\sum_{n=1}^N\phi_{n,t}^{(N)}\\
&\ge \liminf_{N\to\infty}\frac1T\sum_{t=1}^T\frac1N\sum_{n=1}^N\phi_{n,t}^{0}
\ge0,
\end{align*}
where the last inequality follows from $\phi_{n,t}^0\ge0$ for every $n,t$.

\subsection{Proof of Theorem~\ref{thm:S2_sw}}

Recall that, in the switchback design,
\[
\mathrm{Bias}^{SW,(N)}
= -\frac1{NT}\sum_{n=1}^N\sum_{t=1}^{T-1}
 c_n\Big(
\EE(\hat S_{n,t}^{(N)}(\boldsymbol{1})-D_{n,t})^+
-\EE(\hat S_{n,t}^{(N)}(\boldsymbol{0})-D_{n,t})^+
\Big).
\]
For each $n,t$, define
\[
\varphi_{n,t}^{(N)}
:= -c_n\EE(\hat S_{n,t}^{(N)}(\boldsymbol{1})-D_{n,t})^+
+c_n\EE(\hat S_{n,t}^{(N)}(\boldsymbol{0})-D_{n,t})^+
\]
and
\[
\varphi_{n,t}^{0}
:= -c_n\EE(X_{n,t}(1)-D_{n,t})^+
+c_n\EE(X_{n,t}(0)-D_{n,t})^+.
\]
By the convex-order argument in the proof of Proposition~\ref{thm:S2_gte} and $c_n\ge0$, we have $\varphi_{n,t}^{0}\ge0$. This sign is established for the limiting quantities $X_{n,t}(w)$, where the multiplier is deterministic and common across assignments; no sign is claimed for $\varphi_{n,t}^{(N)}$ before taking the mean-field limit.
The Lipschitz argument in~\eqref{eq:s2_global_revenue_average_convergence}, with $R^+$ replaced by $(s-D)^+$ and multiplied by the uniformly bounded $c_n$, gives
\[
\frac1N\sum_{n=1}^N |\varphi_{n,t}^{(N)}-\varphi_{n,t}^{0}|\to 0
\quad\text{for each fixed }t.
\]
Therefore, since $T$ is fixed,
\begin{align*}
\liminf_{N\to\infty}\mathrm{Bias}^{SW,(N)}
&=\liminf_{N\to\infty}\frac1T\sum_{t=1}^{T-1}\frac1N\sum_{n=1}^N\varphi_{n,t}^{(N)}\\
&\ge \liminf_{N\to\infty}\frac1T\sum_{t=1}^{T-1}\frac1N\sum_{n=1}^N\varphi_{n,t}^{0}
\ge0.
\end{align*}

\subsection{Proof of Theorem~\ref{thm:S2_ir}}

In IR, each item is permanently assigned to treatment or control. The bias expression compares, for $w\in\{0,1\}$,
\[
\EE\!\big[ R_{n,t}^+(\hat S_{n,t}^{(N)}(\W_t),D_{n,t})\mid W_{n,t}=w \big]
\quad\text{with}\quad
\EE R_{n,t}^+\big(\hat S_{n,t}^{(N)}(w\boldsymbol 1),D_{n,t}\big).
\]
Let $p_w:=\PP(W_{n,t}=w)$; thus $p_1=p$ and $p_0=1-p$. Using the uniform Lipschitz bounds from Proposition~\ref{thm:S2_gte}, for each fixed $t$ and $w$,
\begin{align*}
&\frac1N\sum_{n=1}^N
\left|
\EE\big[ R_{n,t}^+(\hat S_{n,t}^{(N)}(\W_t),D_{n,t})\mid W_{n,t}=w\big]
-
\EE R_{n,t}^+\big(\hat S_{n,t}^{(N)}(w\boldsymbol 1),D_{n,t}\big)
\right|\\
&\qquad\le
L_RL_S\frac1N\sum_{n=1}^N
\EE\big[|\hat\lambda_t^{(N)}(\W_t)-\hat\lambda_t^{(N)}(w\boldsymbol 1)|\mid W_{n,t}=w\big]\\
&\qquad=
\frac{L_RL_S}{Np_w}\sum_{n=1}^N
\EE\big[|\hat\lambda_t^{(N)}(\W_t)-\hat\lambda_t^{(N)}(w\boldsymbol 1)|\mathbf 1_{\{W_{n,t}=w\}}\big]\\
&\qquad\le
\frac{L_RL_S}{p_w}\EE|\hat\lambda_t^{(N)}(\W_t)-\lambda_t^*|
+L_RL_S\EE|\hat\lambda_t^{(N)}(w\boldsymbol 1)-\lambda_t^*|.
\end{align*}
The right-hand side converges to zero by Lemma~\ref{lem:lambda_limit}, including its deterministic-global-assignment extension. Averaging over the finitely many periods $t$ and using the IR bias expression then yields
\[
\lim_{N\to\infty}\mathrm{Bias}^{IR,(N)}=0.
\]

\subsection{Proof of Theorem~\ref{thm:S2_pr}}

Recall that
\begin{align*}
    & \quad  Bias^{PR,(N)}  - Bias^{SW,(N)} \\
    & =\frac{1}{NT}\sum_{n=1}^N \sum_{t=1}^T\Big(\EE[ R_{n,t}^+(\hat S_{n,t}^{(N)}(\W_{t}),D_{n,t}) \mid W_{n,t} = 1] - \EE R_{n,t}^+(\hat S_{n,t}^{(N)}(\boldsymbol{1}),D_{n,t}) \Big)  \\
       & \quad - \frac{1}{NT}\sum_{n=1}^N \sum_{t=1}^T\Big(\EE[ R_{n,t}^+(\hat S_{n,t}^{(N)}(\W_{t}),D_{n,t}) \mid W_{n,t} = 0] - \EE R_{n,t}^+(\hat S_{n,t}^{(N)}(\boldsymbol{0}),D_{n,t}) \Big)  \\
   & \quad -  \frac{1}{NT} \sum_{n=1}^N \sum_{t=1}^{T-1} c_n \Big(\EE[(\hat S_{n,t}^{(N)}(\W_{t}) - D_{n,t})^+ \mid W_{n,t} = 1]  - \EE[(\hat S_{n,t}^{(N)}(\boldsymbol{1}) - D_{n,t})^+  ] \Big) \\
      & \quad + \frac{1}{NT} \sum_{n=1}^N \sum_{t=1}^{T-1} c_n \Big(\EE[(\hat S_{n,t}^{(N)}(\W_{t}) - D_{n,t})^+ \mid W_{n,t} = 0]  - \EE[(\hat S_{n,t}^{(N)}(\boldsymbol{0}) - D_{n,t})^+  ] \Big).
\end{align*}

Let $p_w:=\PP(W_{n,t}=w)$, so $p_1=p$ and $p_0=1-p$. The first two averaged terms vanish by the same multiplier-averaging argument used in the proof of Theorem~\ref{thm:S2_ir}. For the leftover terms, the same argument applies with the Lipschitz function $(s-D_{n,t})^+$ and the uniformly bounded coefficients $c_n$. Specifically, for $w\in\{0,1\}$,
\begin{align*}
&\frac1N\sum_{n=1}^N c_n
\left|
\EE[(\hat S_{n,t}^{(N)}(\W_t)-D_{n,t})^+\mid W_{n,t}=w]
-
\EE[(\hat S_{n,t}^{(N)}(w\boldsymbol 1)-D_{n,t})^+]
\right|\\
&\qquad\le
\frac{\bar c L_S}{p_w}\EE|\hat\lambda_t^{(N)}(\W_t)-\lambda_t^*|
+\bar c L_S\EE|\hat\lambda_t^{(N)}(w\boldsymbol 1)-\lambda_t^*|
\to 0,
\end{align*}
where $\bar c:=\sup_n c_n<\infty$. Hence
\[
\mathrm{Bias}^{PR,(N)}-\mathrm{Bias}^{SW,(N)}\to 0.
\]
Combining this convergence with Theorem~\ref{thm:S2_sw} gives
\[
\liminf_{N\to\infty}\mathrm{Bias}^{PR,(N)}
= \liminf_{N\to\infty}\mathrm{Bias}^{SW,(N)} \ge 0.
\]
This establishes the theorem without requiring the individual bias sequences to converge.

\section{Implementation Details and Experimental Configuration}
\label{app:implementation_details}

This appendix collects implementation details for the two numerical modules in Section~\ref{sec:numerical}. Section~\ref{app:controlled_stochastic_details} describes the controlled stochastic simulations. Section~\ref{app:freshretail_details} describes the trace-driven FreshRetailNet simulations used for real-world mechanism validation.

\subsection{Controlled Stochastic Simulations}
\label{app:controlled_stochastic_details}

We generate true demand according to the same uniform family used in the theoretical analysis:
\[
D_{n,t}=\mu_n+\alpha_nU_{n,t},\qquad U_{n,t}\sim\mathrm{Unif}[-1,1].
\]
Then we solve the estimated myopic base-stock problem period by period using the estimated uniform parameters $(\hat\mu_{n,t},\hat\alpha_{n,t})$. Capacity is parameterized as
\[
B=\rho\sum_n q_n^0,
\qquad
q_n^0=\mu_n+\alpha_n\left(2\frac{b_n-c_n}{b_n-c_n+h_n}-1\right),
\]
where $\rho$ is the capacity factor.

\textbf{Scenario~1.}
We use $N=3000$, $T=60$, $p=0.5$, 300 global-treatment/global-control replications, and 300 design replications. Forecasts satisfy
\[
\hat\mu_{n,t}(w)=\mu_n+\alpha_n\Delta(w),\qquad \hat\alpha_{n,t}=\alpha_n,
\]
with $\Delta(0)=-0.50$ and $\Delta(1)=-0.05$. Item parameters are generated as
\[
\mu_n\sim\mathrm{Unif}[40,100],\qquad
\alpha_n=\mu_nV_n, \quad V_n\sim\mathrm{Unif}[0.25,0.45],
\]
\[
b_n\sim\mathrm{Unif}[9.5,10.5],\qquad
c_n\sim\mathrm{Unif}[1.8,2.2],\qquad
h_n\sim\mathrm{Unif}[1.2,1.8].
\]
The reported capacity factors are $0.90$, $0.92$, and $1.20$.

\textbf{Scenario~2.}
We use $N=3000$, $T=60$, $p=0.5$, 300 global-treatment/global-control replications, and 300 design replications. Forecasts satisfy
\[
\hat\mu_{n,t}(w)=\mu_n+\epsilon_{n,t}(w),\qquad \hat\alpha_{n,t}=\alpha_n,
\]
where
\[
\epsilon_{n,t}(0)\sim\mathrm{Unif}[-30,30],\qquad
\epsilon_{n,t}(1)\sim\mathrm{Unif}[-0.2,0.2].
\]
Thus both forecasts are centered at the true mean, but treatment has much smaller dispersion. Item parameters are generated as
\[
\mu_n\sim\mathrm{Unif}[100,160],\qquad
\alpha_n\sim\mathrm{Unif}[25,35],
\]
\[
b_n\sim\mathrm{Unif}[9.8,10.2],\qquad
c_n\sim\mathrm{Unif}[1.9,2.1],\qquad
h_n\sim\mathrm{Unif}[1.4,1.6].
\]
The reported capacity factors are $0.85$, $1.00$, and $1.10$.

For both scenarios, the selected regimes pass the numerical validity checks associated with the main regularity assumptions used in the theoretical analysis.

\subsection{FreshRetailNet Trace-Driven Simulations}
\label{app:freshretail_details}

All trace-driven experiments use the local FreshRetailNet-50K split with a 90-day training window and a 7-day evaluation horizon. Forecasting models and the inventory simulator operate on daily targets.

\subsection{Latent Demand Recovery}

Observed sales are censored by stockouts. We therefore use the TimesNet-based recovery component adapted from the FreshRetailNet-50K baseline repository~\citep{wang2025freshretailnet}. The recovered field \texttt{sale\_amount\_pred} is used as the latent daily demand trace in the A/B-test simulator. Forecasting models are trained either on raw censored demand or on TimesNet-recovered demand, depending on the scenario.

\subsection{Forecasting Models}
\label{appendix:forecasting-models}

All forecasting models output point predictions, interpreted as demand-mean forecasts in the trace-driven simulator.

\textbf{Naive.}
The naive benchmark predicts demand on day $t$ using demand from day $t-7$ for the same store--product series.

\textbf{DLinear.}
DLinear uses trend--seasonality decomposition with a linear projection head. The runner uses a 90-day total sequence, input length 62, prediction length 7, moving-average window 28, MAE loss, batch size 1024, learning rate 0.001, and 6 training epochs. RevIN is disabled in the default configuration.

\textbf{Temporal Fusion Transformer (TFT).}
TFT is trained for the 7-day horizon and consumed in deterministic prediction mode. The runner uses maximum encoder length 70, prediction length 7, hidden size 32, attention head size 2, hidden continuous size 16, dropout 0.1, one LSTM layer, learning rate 0.01, batch size 1024, gradient clipping value 0.1, and 5 training epochs. GPU training and prediction are used when available.

\subsection{Synthetic Price and Cost Parameters}

FreshRetailNet-50K does not provide the full economic parameters required by the inventory simulator, so we synthesize prices and costs using a fixed experiment seed. For product $i$, store $s$, and day $t$, the simulator constructs
\[
\mathrm{price}_{i,s,t}=\mathrm{base}_i\cdot\mathrm{category}_i\cdot\mathrm{store}_s\cdot\mathrm{holiday}_t.
\]
The product base factor is sampled as $\mathrm{base}_i\sim\mathcal U[10,90]$, the category factor as $\mathrm{category}_i\sim\mathcal U[0.8,1.2]$, and the store factor as $\mathrm{store}_s\sim\mathcal U[0.9,1.1]$. On holidays, $\mathrm{holiday}_t\sim\mathcal U[0.98,1.02]$; otherwise $\mathrm{holiday}_t=1$. Ordering and holding costs are generated as
\[
\mathrm{ordering\ cost}_{i,s,t}\sim\mathcal U[0.3,0.6]\times \mathrm{price}_{i,s,t},
\qquad
\mathrm{holding\ cost}_{i,s,t}\sim\mathcal U[0,0.3]\times \mathrm{ordering\ cost}_{i,s,t}.
\]
The realized selling price is $\mathrm{price}_{i,s,t}\times\mathrm{discount}_{i,s,t}$, where the discount variable is taken from the dataset when available and set to one otherwise. These synthetic economic parameters are fixed across randomizations within each experiment configuration.

\subsection{Capacity Construction and Inventory Simulation}

Store-level capacity is fixed over the 7-day evaluation horizon. Let $N_s$ be the number of products in store $s$, and let $\tilde d$ denote the median, across store--product series, of the evaluation-period mean daily observed sales. The store capacity is
\[
B_s=\rho N_s\tilde d.
\]
Scenario~1 uses $\rho\in\{0.9,1.2,1.8\}$ for tight, medium, and loose capacity, respectively. Scenario~2 uses $\rho\in\{0.6,0.9,1.8\}$.

For each store-period, the simulator computes inventory targets from the forecast assigned to each store--product--day. Existing inventory carries over from the previous day. If unconstrained targets exceed store capacity, the simulator allocates remaining capacity in descending order of unit margin, where unit margin is selling price minus ordering cost. Realized reward is revenue from fulfilled demand minus ordering and holding costs; remaining inventory receives a terminal salvage credit in the final period.

\subsection{Stockout Substitution}

When stockout substitution is enabled, unmet demand from a stocked-out product is redistributed to other products in the same store. For each store, the simulator constructs a transition matrix from product-hierarchy similarity. Matches at management-group, first-category, second-category, and third-category levels receive weights 1, 2, 4, and 8, respectively; transition probabilities are obtained by applying a softmax over non-self alternatives. For each stocked-out item, rounded unmet demand is redistributed by multinomial draws from this store-specific transition matrix.

We report the substitution ratio
\[
\text{sub-ratio}=\frac{\text{demand received via substitution}}{\text{total effective demand after redistribution}},
\]
averaged across all SKUs. In our trace-driven experiments, this ratio ranges approximately from 19\%--33\% in Scenario~1 and 5\%--20\% in Scenario~2.

\end{document}